\documentclass[a4paper,preprintnumbers,showpacs,onecolumn,superscriptaddress,nofootinbib,amsmath,amssymb,notitlepage]{revtex4-2}

\usepackage{mathtools,leftindex,tensor,mhchem}
\usepackage{booktabs}
\usepackage{siunitx}
\usepackage{enumitem}
\usepackage{times}
\usepackage{dcolumn}
\usepackage{mathrsfs}
\usepackage{comment}
\usepackage{hyperref}
\usepackage{amsmath,accents}
\usepackage{mathtools}
\usepackage{bbm}
\usepackage{bm}
\usepackage{amsfonts}
\usepackage{amssymb}
\usepackage{mathrsfs}
\usepackage[scr=rsfs,cal=boondox]{mathalfa}
\usepackage{wasysym}
\usepackage{graphicx}
\usepackage[]{subfigure}
\usepackage{filecontents}
\usepackage[dvipsnames]{xcolor}
\usepackage{bbold}
\usepackage[scr=rsfs,cal=boondox]{mathalfa}

\usepackage{stackengine}
\stackMath
\newcommand\tsup[2][2]{%
 \def\useanchorwidth{T}%
  \ifnum#1>1%
    \stackon[-1.3ex]{\tsup[\numexpr#1-1\relax]{#2}}{\mathchar"307E}%
  \else%
    \stackon[-1ex]{#2}{\mathchar"307E}%
  \fi%
}
\hypersetup{
    bookmarks=true,         % show bookmarks bar?
    pdftoolbar=true,        % show Acrobats toolbar?
    pdfmenubar=true,        % show Acrobats menu?
    pdffitwindow=true,     %window fit to page when opened
    pdfstartview={FitH},     %fits the width of the page to the window
    pdftitle={My title},     %title
    pdfauthor={author},      %author
    pdfsubject={Subject},    %subject of the document
    pdfcreator={Creator},    %creator of the document
    pdfproducer={Producer},  %producer of the document
    pdfkeywords={keyword1} %{key2} {key3}, % list of keywords
    pdfnewwindow=true,       %links in new window
    colorlinks=true ,        %false: boxed links; true: colored links
    linkcolor=Blue  ,        %color of internal links (change box color with linkbordercolor)
    citecolor=Blue,      %color of links to bibliography
    filecolor=Blue,       %color of file links
    urlcolor=Blue            %color of external links
}

\newcommand{\ed}{\mathrm{d}}

\newcommand{\oalpha}[1]{\accentset{\circ}{\alpha}}
\newcommand{\obf}[1]{\accentset{\circ}{\mathbf{f}}}
\newcommand{\boR}[1]{\accentset{\circ}{\mathbf{R}}}
\newcommand{\obF}[1]{\accentset{\circ}{\mathbf{F}}}
\newcommand{\obPi}[1]{\accentset{\circ}{\mathbf{\Pi}}}

\usepackage{scalerel}
\usepackage{tikz}
\usetikzlibrary{svg.path}

\definecolor{orcidlogocol}{HTML}{A6CE39}
\tikzset{
  orcidlogo/.pic={
    \fill[orcidlogocol] svg{M256,128c0,70.7-57.3,128-128,128C57.3,256,0,198.7,0,128C0,57.3,57.3,0,128,0C198.7,0,256,57.3,256,128z};
    \fill[white] svg{M86.3,186.2H70.9V79.1h15.4v48.4V186.2z}
                 svg{M108.9,79.1h41.6c39.6,0,57,28.3,57,53.6c0,27.5-21.5,53.6-56.8,53.6h-41.8V79.1z M124.3,172.4h24.5c34.9,0,42.9-26.5,42.9-39.7c0-21.5-13.7-39.7-43.7-39.7h-23.7V172.4z}
                 svg{M88.7,56.8c0,5.5-4.5,10.1-10.1,10.1c-5.6,0-10.1-4.6-10.1-10.1c0-5.6,4.5-10.1,10.1-10.1C84.2,46.7,88.7,51.3,88.7,56.8z};
  }
}

\newcommand\orcidicon[1]{\href{https://orcid.org/#1}{\mbox{\scalerel*{
\begin{tikzpicture}[yscale=-1,transform shape]
\pic{orcidlogo};
\end{tikzpicture}
}{|}}}}
\begin{document}

%%%%%%%%%

\title{Null geodesics around a black hole with weakly coupled global monopole charge
%Null geodesics on the spacetime of a weak coupling nonminimal global monopole
%
}

\author{Mohsen Fathi\orcidicon{0000-0002-1602-0722}}
\email{mohsen.fathi@ucentral.cl}
\affiliation{Grupo de Ciencias del Espacio y F\'{i}sicas, Universidad Central de Chile, Toesca 1783, Santiago 8320000, Chile}
%\affiliation{Facultad de Ingenier\'{i}a y Arquitectura, Universidad Central de Chile,  Santa Isabel 1186, 8330563, Santiago, Chile}

\author{J.R. Villanueva\orcidicon{0000-0002-6726-492X}}
\email{jose.villanueva@uv.cl}
\affiliation{Instituto de F\'{i}sica y Astronom\'{i}a, Universidad de Valpara\'{i}so,
Avenida Gran Breta\~{n}a 1111, Valpara\'{i}so, Chile}

\author{Thiago R.P. Caramês\orcidicon{0000-0001-6349-8297}}
\email{trpcarames@id.uff.br}
\affiliation{Departamento de Ciências Exatas, Biológicas e da Terra, Universidade Federal Fluminense (UFF),  Santo Antônio de Pádua, Rio de Janeiro 28470-000, Brazil}

\author{Alejandro Morales-Díaz}
\email{alejandro.moralesd@alumnos.uv.cl}
\affiliation{Instituto de F\'{i}sica y Astronom\'{i}a, Universidad de Valpara\'{i}so,
Avenida Gran Breta\~{n}a 1111, Valpara\'{i}so, Chile}

%\date{\today}

%%%%%%%%%%%
\begin{abstract}

In this paper, we study an asymptotically flat black hole spacetime with weakly nonminimally coupled monopole charge. We analytically and numerically investigate light ray propagation around such a black hole by employing the common Lagrangian formalism. Our analysis encompasses both radial and angular geodesics, for which we present analytical solutions in terms of incomplete Lauricella hypergeometric functions. Additionally, we explore the impact of the coupling constant on geodesic motion. Based on observations from the Event Horizon Telescope, we constrain the black hole parameters, resulting in a coupling constant range of $-0.5 \lesssim \alpha\lesssim 0.5 $. Throughout our analysis, we simulate all possible trajectories and, where necessary, perform numerical inversion of the included integrals.

\bigskip

{\noindent{\textit{keywords}}: Black holes, nonminimal coupling, monopole charge, null geodesics
%gravitation, black holes, numerical methods, astrophysics
}\\

\noindent{PACS numbers}: 04.20.Fy, 04.20.Jb, 04.25.-g   
\end{abstract}

\maketitle

\tableofcontents

%%%%%%%%%%%%%%%%%%%%%%%%%%%%%%%%%%%%%%%%%%%%%%sect.I
\section{Introduction and Motivation}\label{sec:intro}

The potential appearance of cosmic topological defects stands as one of the outstanding predictions of grand unified theories \cite{shellard_vilenkin}. These defects are believed to emerge from phase transitions in the early universe, a process described by the Kibble mechanism, which correlates the type of defect formed to the symmetry group broken during a specific cosmological transition \cite{Kibble_1976}. A notable example is the global monopole, a point-like topological defect arising from the breaking of the group $\rm{SO}(3)$ into $\rm{U}(1)$. The gravitational effects of such a monopole were first explored by Barriola and Vilenkin within the framework of General Relativity (GR) \cite{Barriola-Vilenkin:1989}. Among their results, they demonstrated that the gravitational field of the monopole is represented by a Schwarzschild-like metric featuring an additional "charge", with its magnitude dependent on the energy scale of the symmetry breaking responsible for forming the monopole. The mass term in the solution can be interpreted as either the mass contained within the monopole's core or the mass of a static black hole that has devoured the monopole. Furthermore, they showed that for a negligible core mass, the geometry surrounding the monopole reduces to the Minkowski spacetime with a deficit solid angle. Later, Harari and Lousto, utilizing both analytical and numerical techniques, revealed a striking feature of the global monopole: the core mass is negative, implying the presence of a repulsive gravitational potential around it \cite{Harari-Lousto}.

In modified gravity theories, considerable attention has been given to gravitating global monopole solutions \cite{CRomero,Liu_2009,Carames_2011,Carames_2017,Lambaga_2018,Nascimento_2019,gusmann_scattering_2021,GBgravity-2023}. Recently, Caramês explored a new proposal within the context of nonminimally coupled gravity in Ref. \cite{carames_nonminimal_2023}, whichs revealed that the interaction between matter and the geometric sectors could bestow the monopole with several intriguing properties. In particular, the nonminimal coupling was shown to yield a positive mass for the monopole's core, as well as influence the defect's internal structure by altering the size of the core. Another scenario analyzed was the possibility of a nonminimal global monopole acting as an additional hair of a Schwarzschild black hole (SBH), thus reviving an idea originally proposed by Barriola and Vilenkin. This analysis was carried out in the weak coupling regime, where the nonminimal matter-curvature coupling is treated as a small perturbative parameter. Within this framework, both the geodesic motion of time-like particles and the effects on gravitational light bending were studied in Ref. \cite{carames_nonminimal_2023}, seeking potential observational signatures of the global monopole's presence on astrophysical scales. This line of inquiry is valuable, as it enables us to use experimental bounds from the astrophysical environment to potentially constrain both the strength of the nonminimal coupling and the global monopole parameter, which, in turn, could provide insights into the symmetry-breaking scale associated with the formation of the defect.

Motivated by these results, we aim to further investigate particle dynamics around the black hole solution presented in \cite{carames_nonminimal_2023}, focusing on null geodesics to obtain a more complete description of the spacetime in question.

In this work, we derive exact analytical solutions for light rays propagating around a black hole with a weakly coupled global monopole charge. In fact, the motion of planets and light along geodesics was instrumental in the early success of general relativity. Key tests included the prediction and measurement of light deflection during the 1919 solar eclipse \cite{1920RSPTA.220..291D} and the precise calculation of Mercury's perihelion precession \cite{RevModPhys.19.361}. The non-linear nature of the governing equations often requires simplified numerical approaches to the geodesic equations; thus, analytical solutions are invaluable for validating numerical methods and systematically exploring parameter spaces to predict astrophysical observables. Since Hagihara's seminal 1931 study of geodesics in Schwarzschild spacetime \cite{1930JaJAG...8...67H}, significant efforts have been made to derive exact analytical solutions for massive and massless particle geodesics. Recent attention has focused on using modular forms to solve the (hyper-)elliptic integrals that arise in geodesic studies, building on the work of mathematicians such as Jacobi \cite{jacobi_2013}, Abel \cite{abel_2012}, Riemann \cite{Riemann:1857,Riemann+1866+161+172}, and Weierstrass \cite{Weierstrass+1854+289+306}. Numerous studies have applied hypergeometric, elliptic, and Riemannian theta functions to analyze time-like and null geodesics in static and stationary black hole spacetimes (see, e.g., Refs. \cite{kraniotis_general_2002,kraniotis_compact_2003,kraniotis_precise_2004,kraniotis_frame_2005,cruz_geodesic_2005,kraniotis_periapsis_2007,hackmann_complete_2008,hackmann_geodesic_2008,hackmann_analytic_2009,hackmann_complete_2010,olivares_motion_2011,kraniotis_precise_2011,cruz_geodesic_2013,villanueva_photons_2013,kraniotis_gravitational_2014,soroushfar_analytical_2015,soroushfar_detailed_2016,hoseini_analytic_2016,hoseini_study_2017,fathi_motion_2020,fathi_classical_2020,fathi_gravitational_2021,gonzalez_null_2021,fathi_analytical_2021,kraniotis_gravitational_2021,fathi_study_2022,soroushfar_analytical_2022,battista_geodesic_2022,fathi_spherical_2023,fathi_analytical_2023}).

Continuing along these lines, in this paper, we derive exact analytical solutions to the equations of motion for massless particles traversing the exterior geometry of a monopole black hole. This analysis requires careful treatment of integrals with specific properties, and we employ specialized methods for deriving these solutions. In particular, we utilize Lauricella hypergeometric functions with multiple variables, extending previous work in which such functions were used to solve hyper-elliptic integrals in angular geodesics \cite{fathi_analytical_2023}. For the first time, we demonstrate that Lauricella functions can solve non-elliptic integrals that arise in the calculation of radial and angular geodesics, and we apply these results to simulate possible orbits.

The structure of the paper is as follows: In Sect. \ref{sec:overview}, we review the black hole with a global monopole charge introduced in Ref. \cite{carames_nonminimal_2023}, paying particular attention to the formation of black hole horizons and their dependence on parameter variations. In Sect. \ref{sec:dynamics}, we focus on deriving the equations of motion for massless particles using a Lagrangian formalism. Specifically, we examine radially infalling photons and derive the analytic expressions for the radial evolution of the affine parameter and coordinate time. We then turn to angular geodesics, presenting the relevant effective potential and characterizing possible orbits. The general solution to the angular trajectories is given in terms of a high-order Lauricella hypergeometric function, and we perform numerical inversion of the integrals. We also simulate possible orbits in the equatorial plane and illustrate the effects of the nonminimal coupling. Finally, in Sect. \ref{sec:EHTconst.}, we use the properties of critical orbits to compare the theoretical size of the black hole shadow with observations of M87* and Sgr A* from the Event Horizon Telescope (EHT), and we derive constraints on the black hole parameters. We conclude in Sect. \ref{sec:conclusions}.

Throughout this work, we adopt natural units where $G=c=1=M_{\rm{pl}}$. The sign convention is $(- + + \,+)$, and primes denote differentiation with respect to the radial coordinate.

%%%%%%%%%%%%%%%%%%%%%%%%%%%%%%%%%%%%%%%%%%%%sect.II
\section{Overview on the black hole with global monopole charge in the weak coupling regime }\label{sec:overview}

The generalized theory of gravity concerning  nonminimal matter-curvature coupling is given by the action \cite{Bertolami:2007}
\begin{equation}
\mathcal{S} = \int \ed^4 x \sqrt{-g}\left\{\frac{1}{2}f_1(R)+\Bigl[1+\alpha f_2(R)\mathcal{L}_m\Bigr]\right\},
    \label{eq:action0}
\end{equation}
in which $\mathcal{L}_m$ is the matter Lagrangian density, $f_1(R)$ and $f_2(R)$ are arbitrary functions of the Ricci scalar, and $\alpha$ corresponds to the strength of the interactions between $f_2(R)$ and $\mathcal{L}_m$. In other words, $\alpha$ measures the strength of the nonminimal coupling. In Ref. \cite{carames_nonminimal_2023}, the specific choices $f_1(R)=R/(8\pi)$ and $f_2(R)=R$ were considered. In this case, Ref. \cite{PhysRevD.105.024020} imposed an upper bound of $|\alpha|< 5\times 10^{-12}\,\rm{m}^2$ on the coupling parameter, derived from nuclear physics due to the high densities at nuclear scales \cite{PhysRevD.105.044048}. However, since our study does not involve such environments, we discard this upper bound. It is important to note that, for the above choices, setting $\alpha=0$ directly recovers GR. In Ref. \cite{carames_nonminimal_2023}, the matter Lagrangian density in Eq. \eqref{eq:action0} has been specified to 
\begin{equation}
\mathcal{L}_m = -\frac{1}{2} \partial_\mu \varphi^a \partial^\mu \varphi^a-\frac{1}{4}\lambda\bigl(\varphi^{a}\varphi^{a}-\eta^2\bigr)^2,
    \label{eq:Lm_0}
\end{equation}
corresponding to a symmetry breaking from the $\rm{SO}(3)$ to $\rm{U}(1)$, leading to the formation of the global monopole. In Eq. \eqref{eq:Lm_0}, the parameters $\lambda$ and $\eta$ represent the self-interaction constant of the Higgs field $\varphi^a$, and the energy scale associated with the symmetry breaking, respectively. The Higgs field, itself, is an isotriplet of scalar fields of the form $\varphi^a=\eta h(r) \hat{x}^a$, where $a=1,2,3$ and $x^a=\left\{\sin\theta \cos\phi,\sin\theta\sin\phi,\cos\theta\right\}$, in the usual Schwarzschild coordinates $(t,r,\theta,\phi)$, in which the radius-dependent function $h(r)$ obeys the conditions $h(0)=0$ and $h(\infty)=1$. Hence, based on this symmetry which is obeyed by the global monopole, one can consider the spherically symmetric line element
\begin{equation}
\ed s^2=-B(r)\ed t^2+A(r) \ed r^2+r^2\left(\ed\theta^2+\sin^2\theta\ed\phi^2\right),
    \label{eq:ds_0}
\end{equation}
to analyze the field equations. In Ref. \cite{carames_nonminimal_2023}, the field equations were solved for the aforementioned metric ansatz, resulting in explicit expressions for the lapse function $B(r)$ and the metric function $A(r)$. In this study, however, we consider a special case where the matter-gravity coupling is weak. Specifically, we assume that $\alpha f_2(R) = \alpha R < 1$. In this scenario, small values of $\alpha$ are considered, which, in our weak field context, are defined in terms of a length scale $l_0$. Thus, the condition $\alpha < l_0^2$ is satisfied in this limit. When the global monopole is treated as a charge for a black hole of mass $M$, we can expect $l_0 \sim M$. Within this formalism, Ref. \cite{carames_nonminimal_2023} obtained the following asymptotically flat solutions for the weak matter-gravity coupling:
\begin{eqnarray}
    && A(r)^{-1} \approx 1-\Delta -\frac{2M}{r}+\frac{14\alpha\Delta}{r^2}-\frac{18\alpha\Delta M}{r^3}\equiv \delta-\frac{r_s}{r}\pm \frac{r_c^2}{3\delta \,r^2}\mp \frac{3r_c^2 r_s}{14 \delta\, r^3},\label{eq:A(r)}\\
    && B(r) \approx 1-\Delta-\frac{2M}{r}+\frac{20 \alpha \Delta}{r^2}-\frac{30 \alpha \Delta M}{r^3}\equiv \delta-\frac{r_s}{r}\pm \frac{10 r_c^2}{21\delta \,r^2}\mp \frac{5r_c^2 r_s}{14 \delta\, r^3},\label{eq:B(r)}
\end{eqnarray}
where $\Delta = 8\pi\eta^2$, which corresponds to the deficit solid angle. The upper (lower) sign corresponds to the case $\alpha > 0$ ($\alpha < 0$). Furthermore, we have conveniently defined the quantities $r_s = 2M$, $\delta = 1 - \Delta$, and $r_c = \sqrt{42 |\alpha| (1 - \delta) \delta}$. It can be easily verified that $\mathrm{dim}[\alpha] = \mathrm{length}^2$ and that $\Delta$ is dimensionless, thus $\mathrm{dim}[r_c] = \mathrm{length}$, while $\delta$ is also dimensionless. In the absence of the monopole, where $\Delta = 0$ (which implies $\delta = 1$ and $r_c = 0$), the SBH is recovered. It is important to note that the solutions above satisfy the condition $\eta^2 < 1$, indicating that the symmetry-breaking scale remains well below the Planck scale. \\

\begin{figure}[t]
    \centering
    \includegraphics[width=7 cm]{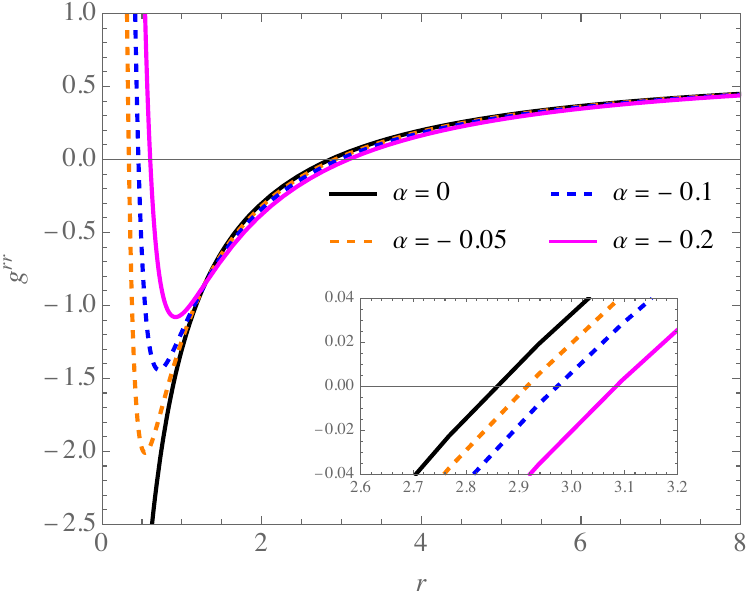} (a)\qquad
    \includegraphics[width=7 cm]{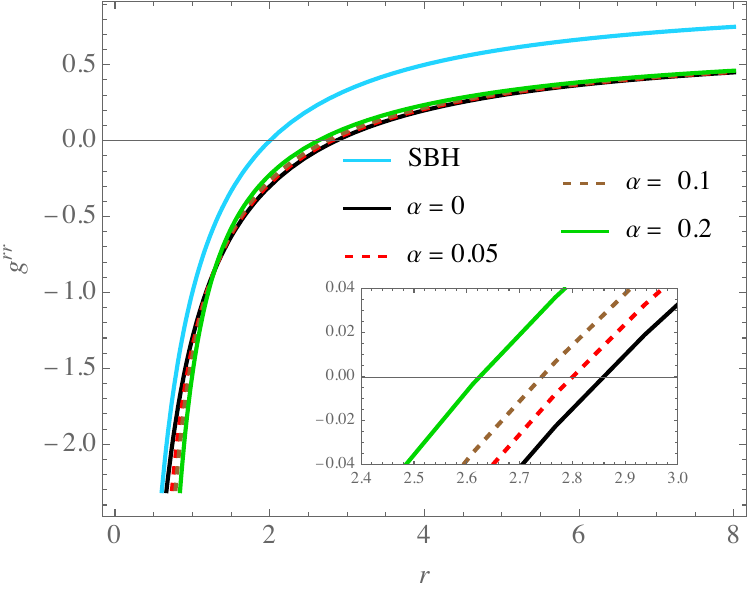} (b)
    \caption{The radial profile of $g^{rr}$ plotted for $\delta=0.7$ The diagrams correspond to (a) $\alpha\leq 0$, and (b) $\alpha\geq0$, including the SBH case with $\delta=1$. The unit of length along the axes is $M$.}
    \label{fig:A(r)-}
\end{figure}

\begin{figure}[h]
    \centering
\includegraphics[width=7 cm]{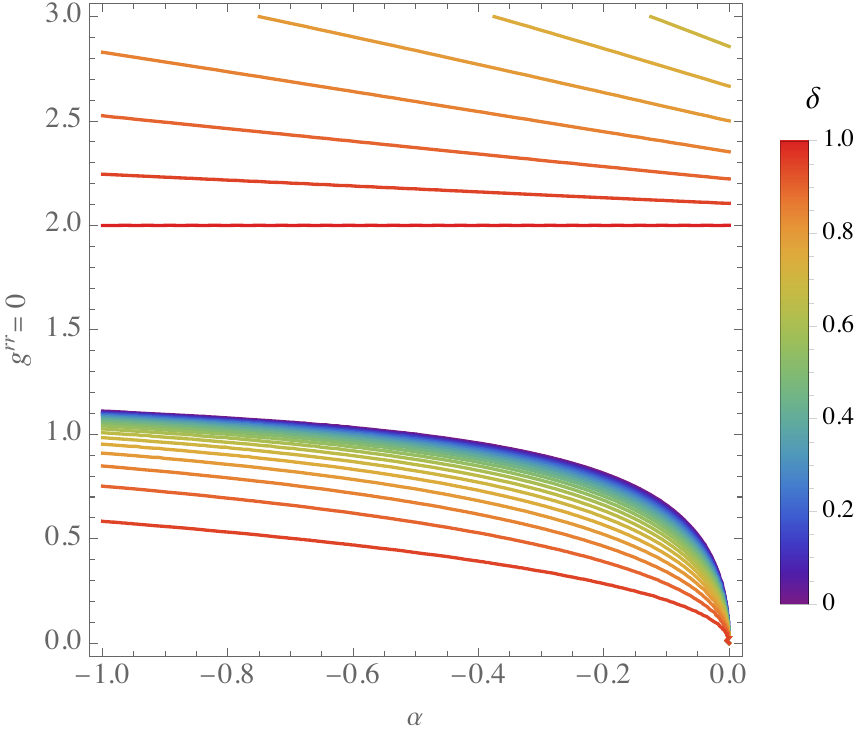} (a)\qquad
\includegraphics[width=7 cm]{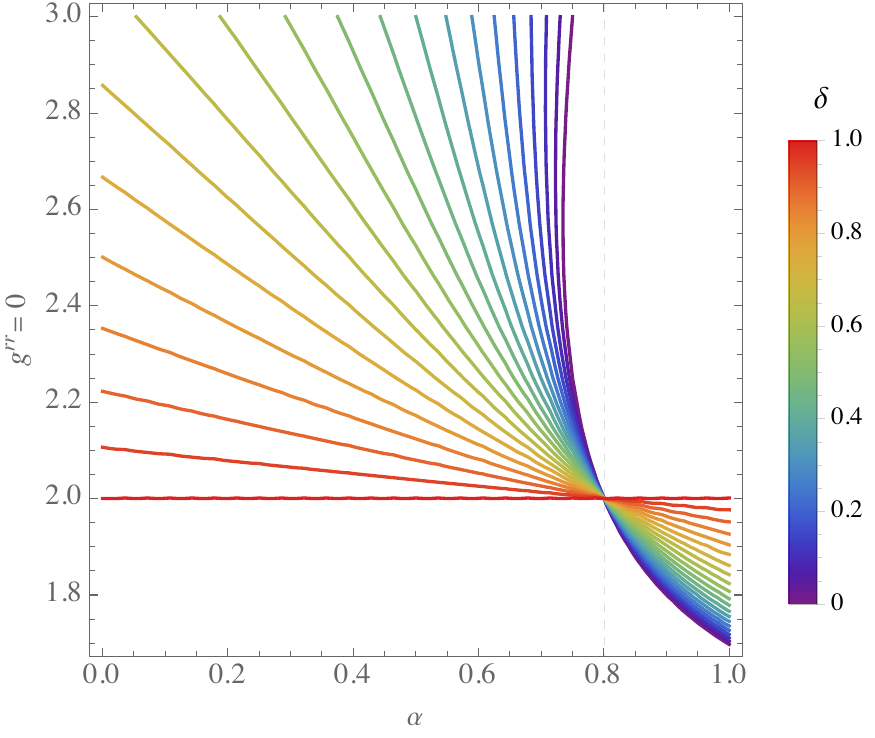} (b)
    \caption{Mutual behavior of $g^{rr}=0$ and $\alpha$ for changes in $\delta$, plotted for (a) $\alpha\leq0$, and (b) $\alpha\geq0$. 
    The dashed vertical line corresponds to the limiting value of $\alpha_*=0.8$. 
    The unit of length along the vertical axes is $M$.}
    \label{fig:rp_alpha}
\end{figure}

The black hole's horizons are where the condition $g^{rr}=0$ is satisfied. This equation has the three analytical solutions
\begin{eqnarray}
    && r_1 = \frac{r_s}{3 \delta}
    \left[1-\sqrt{1\mp \left(\frac{r_c}{r_s}\right)^2}\left(\cos\varphi_0
    -\sqrt{3}\sin \varphi_0\right)
    \right],\label{eq:r1}\\
    && r_2 =  \frac{r_s}{3 \delta}
    \left[1+2\sqrt{1\mp \left(\frac{r_c}{r_s}\right)^2}\,\cos\varphi_0
    \right],\label{eq:r2}\\
    && r_3 = \frac{r_s}{3 \delta}
    \left[1-\sqrt{1\mp \left(\frac{r_c}{r_s}\right)^2}\left(\cos\varphi_0
    +\sqrt{3}\sin \varphi_0\right)
    \right],\label{eq:r3}
\end{eqnarray}
in which (with $\delta_c=14/27\approx 0.519$ )
\begin{equation}
    \label{angroot}
    \varphi_0=\frac{1}{3}\arccos \Biggl(\left| 1\mp \frac{3 r_c^2}{2 r_s^2}
    \left[1-\frac{\delta}{\delta_c}\right]\right|
    \frac{r_s^3}{(r_s^2\mp r_c^2)^{3/2}}\Biggr).
\end{equation}
Accordingly,  we can recast Eq. \eqref{eq:A(r)} as 
\begin{equation}
A(r)^{-1} = \frac{\delta}{r^3}(r-r_1)(r-r_2)(r-r_3).
    \label{eq:A(r)_1}
\end{equation}
 In Fig. \ref{fig:A(r)-}, we have plotted the behavior of $g^{rr}=A(r)^{-1}$, for different positive and negative values of the $\alpha$-parameter, and for a fixed $\delta$. 
As inferred from the diagrams, there are two positive solutions to $g^{rr} = 0$ for the case of $\alpha < 0$ (Fig. \ref{fig:A(r)-}(a)). Thus, the black hole has two horizons. In this case, the solutions in Eqs. \eqref{eq:r1}--\eqref{eq:r3} are characterized as $0 < r_1 < r_2$ and $r_3 < 0$. Therefore, $r_1$ can be identified as $r_-$, corresponding to the Cauchy horizon, while $r_2 = r_+$ represents the event horizon, where the coordinate time becomes null. For $r > r_+$, observers remain time-like, whereas in the region $r_- < r < r_+$, no time-like observers can exist. Moreover, in the domain $0 < r < r_-$, the coordinate time becomes time-like. For the case of $\alpha = 0$, only one horizon exists, as the Cauchy horizon vanishes (since $r_1 = -r_3 \rightarrow 0$). In this scenario, the single horizon is simply given by $r_+ = r_s/(1 - \delta)$. For $\alpha \geq 0$ (Fig. \ref{fig:A(r)-}(b)), the behavior remains consistent with the case of $\alpha = 0$, and the black hole possesses only one horizon, as the equation $g^{rr} = 0$ has only one real solution, $r_2 = r_+$, along with a complex conjugate pair $r_1 = r_3^{*}$. It is evident that the impact of nonminimal coupling in this scenario is minimal, as the profiles are quite similar. However, the contribution of the topological defect encoded in the coefficient $\delta$ is significant, as evidenced by the deviation of the SBH profile from the other curves. To illustrate the general behavior of the event horizon, Fig. \ref{fig:rp_alpha} depicts the changes in solutions to $g^{rr} = 0$ versus variations in the $\alpha$-parameter for various values of $\delta$.
It can be inferred from the diagrams that, in accordance with what we observed in Fig. \ref{fig:A(r)-}(a), for $\alpha \leq 0$, the SBH with $r_+ = r_s$ (for $\delta = 1$) forms the lower bound. As $\delta$ decreases from its maximum value, the black hole size increases. Conversely, the Cauchy horizon shrinks with an increase in the $\alpha$-parameter and ultimately vanishes as $\alpha \rightarrow 0$. For the case of $\alpha \geq 0$, the behavior of $r_+$ is such that the black hole is larger than the SBH for $\alpha < \alpha_*$, where $\alpha_*$ is a critical value beyond which the black hole transits from being larger than the SBH to being smaller than it. This critical value is determined to be $\alpha_* = 0.8$, corresponding to the value of $\alpha$ for which, in Eq. \eqref{eq:A(r)}, we have $A(r_s)^{-1} = 0$. It is important to note that although the internal causal structure of the black hole presents interesting features, for the purpose of this paper, we restrict ourselves to the domain of outer communications, corresponding to $r > r_+$, for all considered values of $\alpha$. Now to further explore the spacetime structure of the black hole, it is essential to identify the hypersurface at which the gravitational redshift becomes infinite. In the spacetime defined by Eqs. \eqref{eq:A(r)} and \eqref{eq:B(r)}, it is evident that this hypersurface does not coincide with that of the event horizon. In fact, the surface of infinite redshift is determined by the equation $g_{tt} = 0$, which corresponds to $B(r) = 0$. This equation is cubic and has the three solutions $r_4$, $r_5$, and $r_6$, which are given by
\begin{eqnarray}
    && r_4 = \frac{r_s}{3 \delta}
    \left[1-\sqrt{1\mp \frac{10}{7}\left(\frac{r_c}{r_s}\right)^2}\left(\cos\overline{\varphi}_0
    -\sqrt{3}\sin \overline{\varphi}_0\right)
    \right],\label{eq:r4}\\
    && r_5 =  \frac{r_s}{3 \delta}
    \left[1+2\sqrt{1\mp \frac{10}{7}\left(\frac{r_c}{r_s}\right)^2}\,\cos\overline{\varphi}_0
    \right],\label{eq:r5}\\
    && r_6 = \frac{r_s}{3 \delta}
    \left[1-\sqrt{1\mp \frac{10}{7}\left(\frac{r_c}{r_s}\right)^2}\left(\cos\overline{\varphi}_0
    +\sqrt{3}\sin \overline{\varphi}_0\right)
    \right],\label{eq:r6}
\end{eqnarray}
where
\begin{equation}
    \label{angroot2}
    \overline{\varphi}_0=\frac{1}{3}\arccos \Biggl(\left| 1\mp \frac{15 r_c^2}{7 r_s^2}
    \left[1-\frac{9\delta}{4}\right]\right|
    \frac{r_s^3}{(r_s^2\mp \frac{10}{7}r_c^2)^{3/2}}\Biggr).
\end{equation}
The above solutions obey the conditions \(0<r_4<r_5\) and \(r_6<0\) for \(\alpha<0\), and \(r_4=r_6^*\) and \(r_5>0\) for \(\alpha>0\). In this regard, for the case of \(\alpha<0\), we assign \(r^{\rm{inf}}_{\rm{in}}=r_4\) and \(r^{\rm{inf}}_{\rm{out}}=r_5\), respectively, as the interior and exterior surfaces of the infinite redshift. On the other hand, for \(\alpha>0\), there is only one such surface, identified by \(r_5\). Finally, we can recast the lapse function as
\begin{equation}
B(r)=\frac{\delta}{r^3}(r-r_4)(r-r_5)(r-r_6).
    \label{eq:B(r)_1}
\end{equation}

%%%%%%%%%%%%%%%%%%%%%%sect.III
\section{Lagrangian formalism for particle dynamics and null geodesics}\label{sec:dynamics}

In the spacetime described by the line element \eqref{eq:ds_0}, the geodesics can be obtained by means of the Lagrangian
\begin{eqnarray}
\mathcal{L}&=&\frac{1}{2}g_{\mu\nu}\dot{x}^{\mu} \dot{x}^{\nu}\nonumber\\
&=& \frac{1}{2}\Bigl[
-B(r) \dot{t}^2+A(r) \dot{r}^2+r^2\dot{\theta}^2+r^2\sin^2\theta\dot{\phi}^2
\Bigr]\nonumber\\
&=&-\frac{1}{2}\epsilon,
    \label{eq:Lagrangian}
\end{eqnarray}
where overdot stands for differentiation with respect to the affine curve parameter, $\tau$. In this context, null and time-like geodesics are characterized by $\epsilon=0$ and $\epsilon=1$, respectively. Defining the generalized momenta $\Pi_\mu=\partial \mathcal{L}/\partial \dot{x}^\mu$, we can establish the two constants of motion
\begin{eqnarray}
    && \Pi_t=-B(r) \dot{t} \equiv -E, \label{eq:E} \\
    && \Pi_\phi= r^2\sin^2\theta\dot{\phi} \equiv L. \label{eq:L}
    \label{eq:conjMoment}
\end{eqnarray}
These constants represent the energy and angular momentum per unit mass associated with the test particles. For the sake of convenience, we confine the geodesics to the equatorial plane by fixing $\theta=\pi/2$, allowing us to write
\begin{equation}
{A(r)B(r)}\dot r^2={E^2}-V(r),
    \label{eq:rdot_gen}
\end{equation}
in which 
\begin{equation}
V(r) = B(r)\left(\epsilon+\frac{L^2}{r^2}\right),
    \label{eq:V(r)_0}
\end{equation}
is the gravitational effective potential. Accordingly, the radial and angular equations of motion are given by
\begin{eqnarray}
    && \left(\frac{\ed r}{\ed t}\right)^2=\frac{B(r)}{A(r)}\left[1-\frac{V(r)}{E^2}\right],\label{eq:drdt_0}\\
    && \left(\frac{\ed r}{\ed \phi}\right)^2 = \frac{r^4}{b^2 A(r) B(r)}\left[1-\frac{V(r)}{E^2}\right],\label{eq:drdphi_0}
\end{eqnarray}
in which, $b=L/E$ is the impact parameter associated with the test particles. These equations are analyzed in the context of radial and angular motions for null trajectories in the remainder of this section.

%%%%%%%%%%%%%%%%%%%
\subsection{Radial null geodesics}\label{subsec:radial}

\begin{figure}
    \centering
    \includegraphics[width=7 cm]{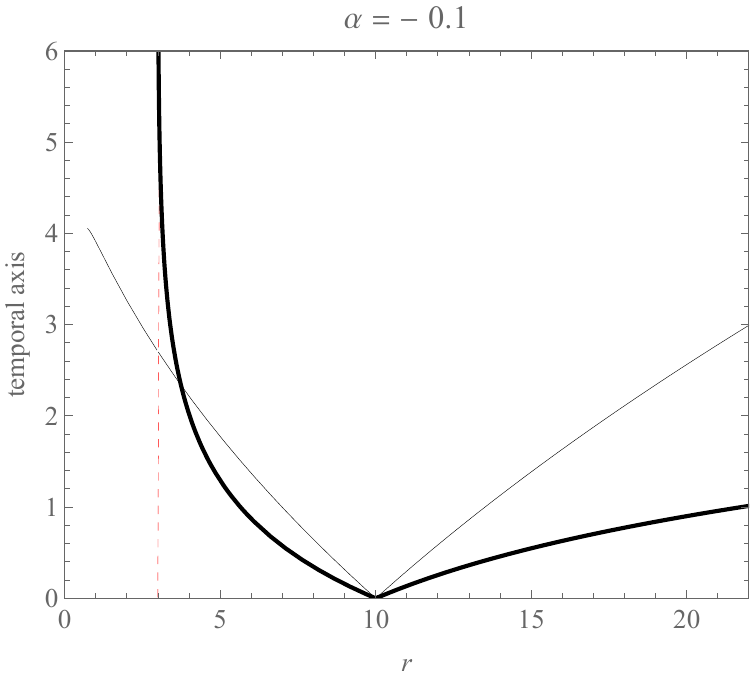} (a)\qquad
    \includegraphics[width=7 cm]{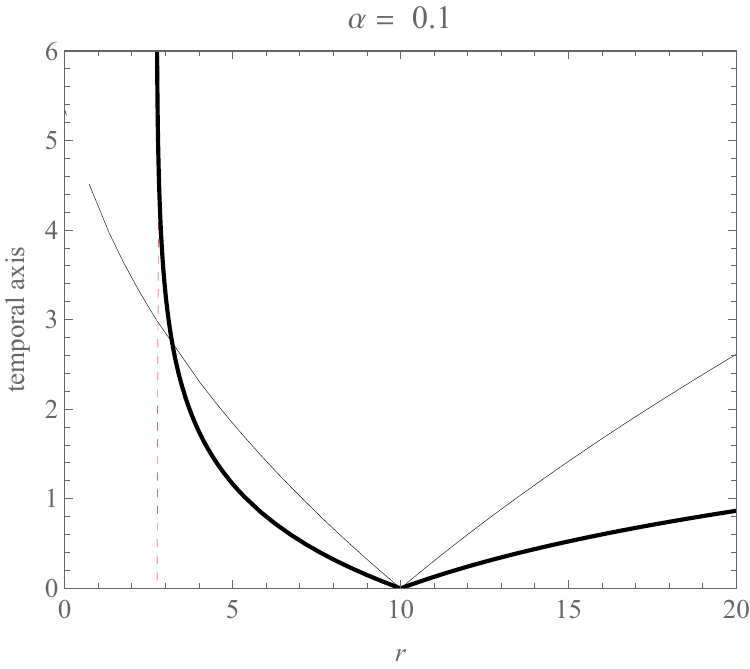} (b)
    \caption{The ingoing and outgoing radial geodesics are plotted as the radial profiles of $\tau(r)$ (thin curves) and $t(r)$ (thick curves), assuming $\delta=0.7$, $r_i=10$, and two different values of the $\alpha$-parameter. The dashed vertical line corresponds to the relevant $r_+$. The unit of length along the axes is $M$.}
    \label{fig:radial}
\end{figure}

Null geodesics are characterized by the condition $\epsilon=0$, leading to the corresponding effective potential given by $V(r) = {L^2 B(r)}/{r^2}$. Furthermore, for purely radial geodesics, where $L=0$ (i.e., $V(r)=0$), the equations of motion are obtained from Eqs. \eqref{eq:rdot_gen} and \eqref{eq:drdt_0} as
\begin{eqnarray}
    && \dot r=\pm\frac{E}{\sqrt{A(r) B(r)}}, \label{eq:rdot_r}\\
    && \frac{\ed r}{\ed t} = \pm\sqrt{\frac{B(r)}{A(r)}},
    \label{eq:drdt_1}
\end{eqnarray}
where the $+\,(-)$ sign corresponds to the outgoing (ingoing) trajectories. By choosing the initial point of approach at $r=r_i$ with $t=\tau=0$, and in accordance with the expressions \eqref{eq:A(r)_1} and \eqref{eq:B(r)_1}, the solutions to the above equations can be obtained through direct integration. For the affine parameter, integrating Eq. \eqref{eq:rdot_r} yields
\begin{equation}
 \tau(u) = \frac{2r_i\sqrt{\tsup[1]{\ell}_5 u^3}}{3E} \,F_D^{(6)}\biggl(
 \frac{3}{2},\frac{1}{2}\frac{1}{2},\frac{1}{2},-\frac{1}{2},-\frac{1}{2},0;\frac{5}{2};c_1,c_2,c_3,c_4,c_6,u
 \biggr), 
 \label{eq:tau(r)}
\end{equation}
in which
\begin{subequations}
    \begin{align}
        &\tsup[1]{\ell}_5=\frac{r_i(r_5-r_4)(r_5-r_6)}{(r_5-r_1)(r_5-r_2)(r_5-r_3)},\\
        & u=1-\frac{r}{r_i},\label{eq:u}\\
        & c_j=\frac{r_i}{r_i-r_j},\quad j=1,2,3,4,6,\label{eq:cj}
    \end{align}
\end{subequations}
and $F_D^{(n+1)}$ is the incomplete $n$-parameter Lauricella hypergeometric function\footnote{See Refs. \cite{fathi_study_2022,fathi_analytical_2023} for additional information and derivation methods.}, which is given in terms of the Euler-type integral
\begin{equation}
\int_0^{z} z^{a-1}(1-z)^{c-a-1}\prod_{i=1}^{n}(1-\xi_i z)^{-b_i}\,\ed z = 
\frac{z^{a}}{a}F_D^{(n+1)}\left(
a,b_1,\dots,b_n,1+a-c;a+1;\xi_1,\dots,\xi_n,z
\right).
    \label{eq:FDn_def}
\end{equation}
Applying the same procedure to the radial evolution of the coordinate time in Eq. \eqref{eq:drdt_0} yields
\begin{equation}
t(u)=\frac{2r_i}{\delta} \sqrt{\frac{u}{\tsup{\ell}_5}} \, 
F_D^{(6)}\biggl(
\frac{1}{2},\frac{1}{2},\frac{1}{2},\frac{1}{2},\frac{1}{2},\frac{1}{2},3;\frac{3}{2};c_1,c_2,c_3,c_4,c_6,u
\biggr),
    \label{eq:t(r)}
\end{equation}
wheres
\begin{equation}
    \tsup{\ell}_5=\frac{1}{r_i^5}(r_5-r_1)(r_5-r_2)(r_5-r_3)(r_5-r_4)(r_5-r_6),
\end{equation}
and the other parameters are the same as those defined in Eqs. \eqref{eq:u} and \eqref{eq:cj}. Based on the solutions obtained above, we have plotted the radial evolution of the affine parameter and coordinate time in Fig. \ref{fig:radial}.
According to the diagrams, we can infer that while the comoving observers perceive themselves passing the horizon within a finite affine parameter, for distant observers, it takes an infinite amount of time for the photons to cross the horizon, making them appear frozen. In this context, the behavior aligns with that of standard spacetimes, such as the Schwarzschild solution and similar configurations \cite{Misner:1973,Chandrasekhar:2002,ryder_2009}.

%%%%%%%%%%%%%%%%%%%%%%%%%%%%%%
\subsection{Angular null geodesics}\label{subsec:angular}

\begin{figure}
    \centering
    \includegraphics[width=7 cm]{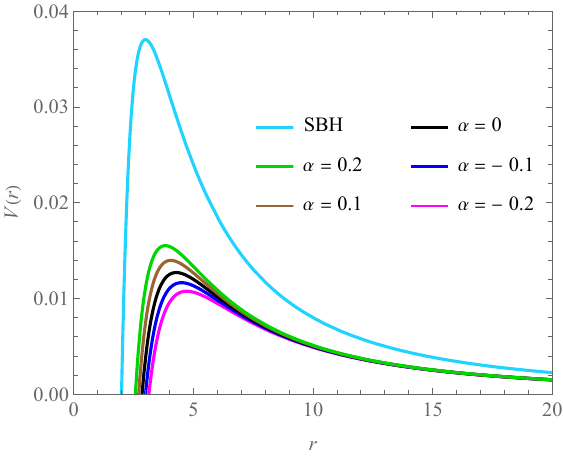} (a)\qquad
    \includegraphics[width=7 cm]{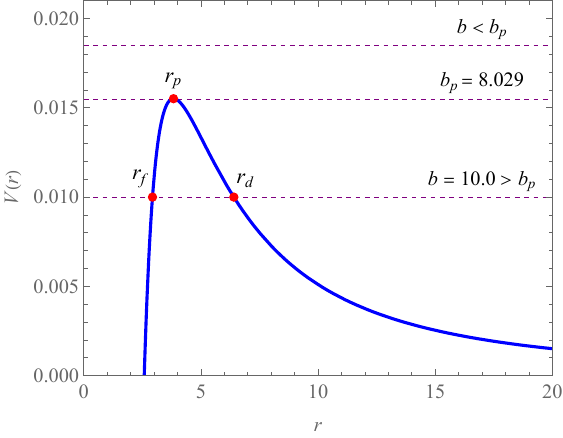} (b)
    \caption{The radial profile of the effective potential is plotted for $L=1$ and $\delta=0.7$. In panel (a), different values of the $\alpha$-parameter are taken into account, and the profiles are compared with the SBH. In panel (b), a typical profile is plotted for $\alpha=0.2$, indicating the turning points corresponding to different types of orbits and their associated impact parameters. For the critical orbits, $r_p=3.829$, and for the OFK and OSK, they are $r_d=6.404$ and $r_f=2.931$, respectively. The unit of length along the radial axes is $M$.}
    \label{fig:Veff}
\end{figure}

For the case of $L\neq0$, the angular equation of motion \eqref{eq:drdphi_0} yields
\begin{equation}
\left(\frac{\ed r}{\ed\phi}\right)^2 = \frac{\mathcal{P}_8(r)}{\mathcal{P}_4(r)},
    \label{eq:drdphi_1}
\end{equation}
where $\mathcal{P}_8(r)=\sum_{j=0}^{8} a_j r^j $ and $\mathcal{P}_4(r)=\sum_{j=1}^{4} \bar{a}_j r^j$, in which
\begin{subequations}
    \begin{align}
       & a_0 =-135 r_s^2 \alpha ^2 b^2 (1-\delta)^2,\\
       & a_1 = 390 r_s \alpha ^2 b^2 (1-\delta )^2,\\
       & a_2 = -280 \alpha ^2 b^2 (1-\delta )^2-24 r_s^2 \alpha  b^2 (1-\delta ),\\
       & a_3 = 34 r_s \alpha  b^2 (1-\delta )+24 r_s \alpha  b^2 (1-\delta ) \delta,\\
       & a_4 = -34 \alpha  b^2 (1-\delta ) \delta -r_s^2 b^2,\\
       & a_5 = 2 r_s b^2 \delta-9 r_s \alpha  (1-\delta ),\\
       & a_6 = 14 \alpha  (1-\delta )-b^2 \delta ^2,\\
       & a_7 = -r_s,\\
       & a_8 = \delta,
    \end{align}
\end{subequations}
and
\begin{subequations}
    \begin{align}
        & \bar{a}_1 = -15 r_s \alpha  b^2 (1-\delta ) ,\\
        & \bar{a}_2 = 20 \alpha  b^2 (1-\delta),\\
        & \bar{a}_3 = 15 r_s \alpha  b^2 (1-\delta),\\
        & \bar{a}_4 = b^2\delta.
    \end{align}
\end{subequations}
Photons approaching the black hole will have zero velocity at points $r_0$, where $\mathcal{P}_8(r_0) = 0$. These roots, therefore, represent the turning points, $r_0 = r_t$, on the gravitational effective potential, where the types of photon motion can be identified. In Fig. \ref{fig:Veff}, we have plotted the radial profile of the effective potential.
As observed in Fig. \ref{fig:Veff}(a), the SBH profile encompasses all other potentials, exhibiting a significantly larger maximum. As the $\alpha$-parameter decreases from positive to negative values, the peak of the effective potential lowers. Thus, a positive nonminimal coupling results in a higher potential barrier for photon motion. In Fig. \ref{fig:Veff}(b), a typical profile is plotted, categorizing possible photon motions based on the photons' initial impact parameters. Each specific impact parameter may or may not encounter turning points, which are identified by the equation $V(r_t)=1/b^2$. In this manner, photon orbits can be classified as follows:

\begin{itemize}
    \item \underline{\textit{Deflecting trajectories}}: For $b > b_p$, photons interact with the potential barrier in two different ways. According to Fig. \ref{fig:Veff}(b), if photons encounter the effective potential at $r = r_d$, they are deflected away from the black hole, escaping the gravitational potential. This type of motion is known as the orbit of the first kind (OFK). Conversely, if the encounter occurs at $r = r_f$, the photon will inevitably fall into the event horizon. This type of orbit is referred to as the orbit of the second kind (OSK).

    \item \underline{\textit{Critical orbits}}: At the potential's maximum, the motion of photons becomes unstable, as they tend to either escape from the black hole or fall onto the event horizon. These orbits are characterized by the impact parameter $b = b_p$ and the radius of critical orbits $r = r_p$ (see Fig. \ref{fig:Veff}(b)). Depending on the final outcome of the trajectories, these orbits are referred to as critical orbits of the first kind (COFK) or the second kind (COSK).

    \item \underline{\textit{Captured orbits}}: For $b < b_p$, the photons approach the black hole with energy higher than the potential's maximum. Consequently, they encounter no turning points and fall directly onto the event horizon.
    
\end{itemize}
We now proceed to analyze these orbits separately and in detail.

%%%%%%%%%%%%%%%%%%%%%%%
\subsubsection{Radius of unstable circular orbits}

\begin{figure}[h]
    \centering
     \includegraphics[width=7 cm]{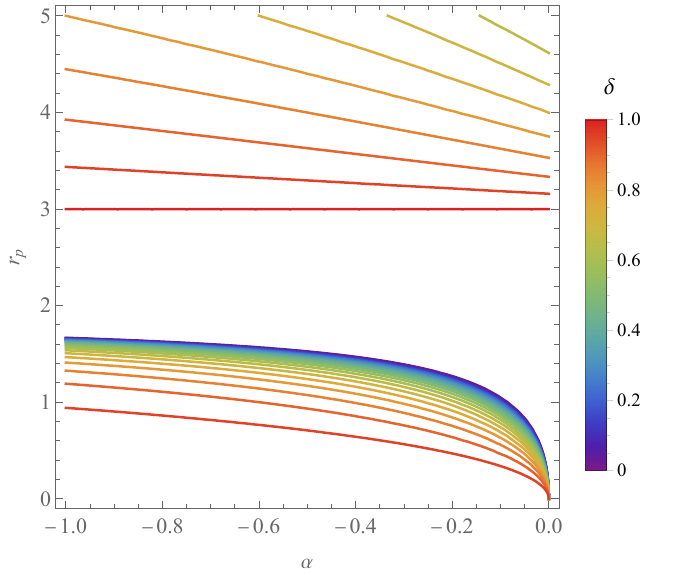} (a)\qquad
     \includegraphics[width=7 cm]{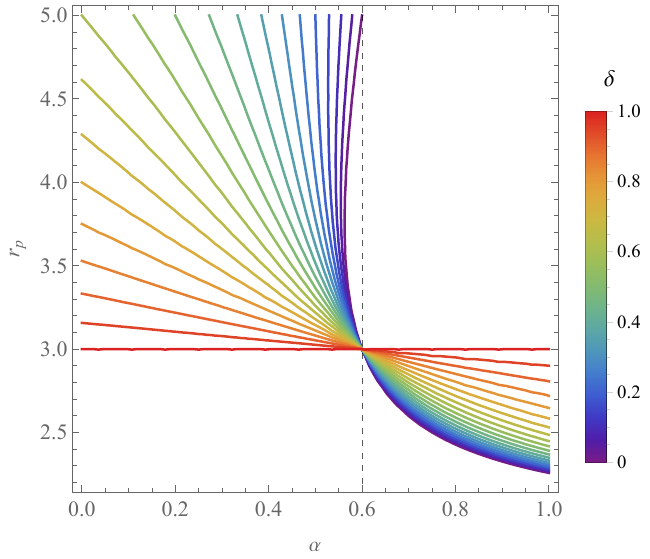} (b)
    \caption{The mutual behavior of $r_p$ and $\alpha$ for changes in $\delta$, plotted for (a) $\alpha\leq0$ and (b) $\alpha\geq0$. The dashed vertical line corresponds to the limiting value of $\alpha_{\rm{SBH}}=0.6$. The unit of length along the vertical axes is $M$.}
    \label{fig:rphoton_alpha}
\end{figure}

The radius of unstable orbits can be determined by solving the equation $V'(r_p)=0$, which has the analytical solution 
\begin{equation}
r_p = -\frac{2}{\delta}\sqrt{\frac{\chi_2}{3}}\cos\left(
\frac{1}{3}\arccos\left(
\frac{3\chi_3}{\chi_2}\sqrt{\frac{3}{\chi_2}}\,
\right)
-\frac{4\pi}{3}
\right)+\frac{r_s}{2\delta},
    \label{eq:rp}
\end{equation}
in which
\begin{subequations}
    \begin{align}
        & \chi_2 =\frac{3r_s^2}{4}-40 \alpha  (1-\delta) \delta,\\
        & \chi_3 = \frac{5}{4} \alpha  \delta  \left(15 \delta ^2-23 \delta +8\right) r_s-\frac{r_s^3}{8}.
    \end{align}
\end{subequations}
It is straightforward to verify that for $\delta = 1$, we obtain $r_p = 2M=3 r_s/2$, which corresponds to the radius of unstable circular orbits (or the photon sphere's radius) for the SBH. In Fig. \ref{fig:rphoton_alpha}, the variation of $r_p$ with respect to the $\alpha$-parameter is shown for both the negative and positive domains of this parameter.
We observe that for $\alpha_{\rm{SBH}} = 0.6$, the radius of the photon sphere transits from being larger than that of the SBH to smaller than that of the SBH. This value is obtained by solving the equation $V'(r)|_{{3r_s}/{2}}=0$ for $\alpha$. Furthermore, the critical value of the impact parameter can be calculated by solving the equation $V(r_p)=1/b_p^2$, which yields
\begin{equation}
b_p = \frac{r_p^{5/2}}{\sqrt{\delta  r_p^3 + 20 \alpha  (1-\delta) r_p - 15 \alpha  (1-\delta) r_s - r_s r_p^2}},
    \label{eq:bp}
\end{equation}
correctly reducing to $b_p=3\sqrt{3}$ for the case of $\delta=1$, corresponding to that for the SBH. To compute the affine and coordinate periods of orbits at $r_p$, we consider the expressions in Eqs. \eqref{eq:E} and \eqref{eq:L}, from which, for a complete revolution, we infer
\begin{eqnarray}
    && T_\tau =  \frac{2\pi r_p^2}{b_p},\label{eq:Ttau}\\
    && T_t =  {2\pi}{b_p}.\label{eq:Tt}
\end{eqnarray}

%%%%%%%%%%%%%%%%%%%%%%
\subsubsection{Deflecting trajectories}

We recall that the polynomials in Eq. \eqref{eq:drdphi_1} can be recast as $\mathcal{P}_8(r)=a_8\prod_{j=1}^{8}(r-\tsup[1]{r}_j)$ and $\mathcal{P}_4(r)=\bar{a}_4r\prod_{j=1}^{3}(r-\tsup{r}_j)$, where $\tsup[1]{r}_{\{\overline{1,8}\}}$ and $\tsup{r}_{\{\overline{1,3}\}}$ are the roots of these polynomials, respectively. Based on the impact parameter specific to the approaching photons, some of the roots $\tsup[1]{r}_j$ may be real and positive. For the case of the deflecting trajectories, as inferred from Fig. \ref{fig:Veff}(b), there are two such roots, $\tsup[1]{r}_7=r_f$ and $\tsup[1]{r}_8=r_d$, which constitute the turning points. The remaining roots include negative values and complex conjugate pairs. 

\begin{figure}[h]
    \centering
    \includegraphics[width=7cm]{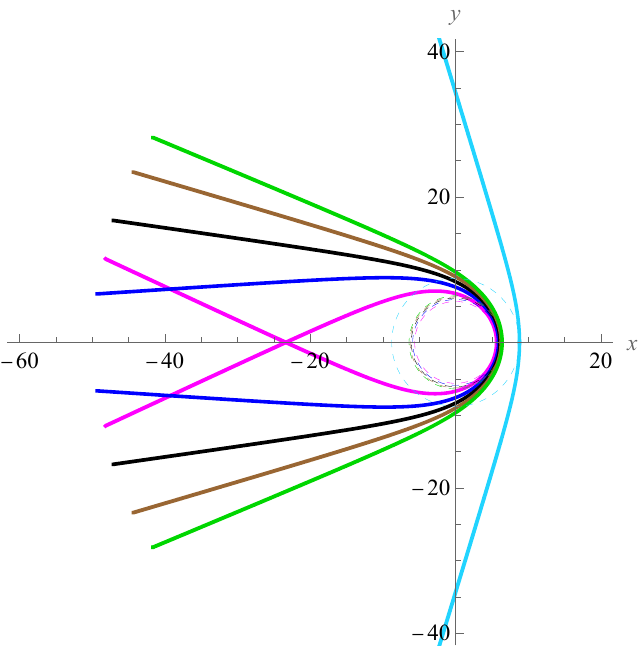} (a)\qquad
    \includegraphics[width=7cm]{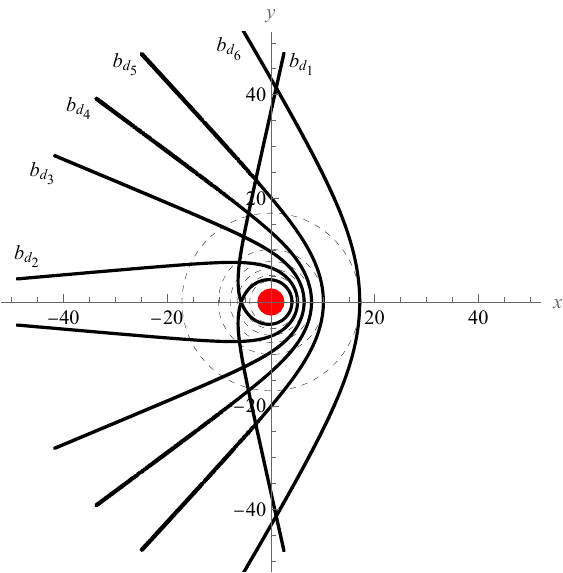} (b)
    \caption{The OFK is plotted for $\phi_0=0$ and $\delta=0.7$. In panel (a), a fixed impact parameter $b_d=10$ is taken into account, with each curve corresponding to a different value of the $\alpha$-parameter. The color-coding is consistent with that in Fig. \ref{fig:Veff}(a). The dashed circles indicate the turning point $r_d$ for each of the orbits, matching the color of the respective curves. Panel (b) shows the OFK for a fixed value of $\alpha=0.2$, in accordance with the configuration in Fig. \ref{fig:Veff}(b), and different impact parameters, specifically $b_{d_1} = b_p+0.04=8.069$, $b_{d_2}=8.771$, $b_{d_3}=10.0$, $b_{d_4}=11.547$, $b_{d_5}=14.142$, and $b_{d_6}=22.361$. The red disk at the center represents the radius of $r_+$. 
    The unit of length along the axes is $M$.}
    \label{fig:OFK}
\end{figure}

For the OFK, by taking into account the above expressions of the polynomials and letting $\phi_0$ be the initial azimuth angle of the approaching photons at the initial point $r_d$, one can resolve the equation of motion \eqref{eq:drdphi_1} to obtain
\begin{equation}
\phi-\phi_0=2b^2\sqrt{X \tilde{\ell}_{-4}} \,F_D^{(11)}\left(
\frac{1}{2},\Bigl\{b_{\{\overline{1,3}\}}\Bigr\},
\Bigl\{b_{\{\overline{4,10}\}}\Bigr\},-\frac{1}{2};\frac{3}{2};
\Bigl\{\tilde{c}_{\{\overline{1,3}\}}\Bigr\},
\Bigl\{\tilde{c}_{\{\overline{4,10}\}}\Bigr\},
X
\right),
    \label{eq:phi(r)_OFK}
\end{equation}
in which
\begin{subequations}
    \begin{align}
       & \Bigl\{b_{\{\overline{1,3}\}}\Bigr\} = \left\{-\frac{1}{2}\right\}_3,\qquad 
        \Bigl\{b_{\{\overline{4,10}\}}\Bigr\} = \left\{\frac{1}{2}\right\}_7,\label{eq:b13410}\\
      & \Bigl\{\tilde{c}_{\{\overline{1,3}\}}\Bigr\} = \left\{\frac{r_d}{r_d-\tsup{r}_{\{\overline{1,3}\}}}\right\}_3,\qquad \Bigl\{\tilde{c}_{\{\overline{4,10}\}}\Bigr\} = \left\{\frac{r_d}{r_d-\tsup[1]{r}_{\{\overline{1,7}\}}}\right\}_7,\\
      & X = 1-\frac{r}{r_d},
    \end{align}
\end{subequations}
where the numbers in the subscripts indicate the number of terms that will be essentially generated by the braces\footnote{For example, $\left\{-\frac{1}{2}\right\}_3=-\frac{1}{2},-\frac{1}{2},-\frac{1}{2}$, and so on.}, and $\tilde{\ell}_{-4}=\prod_{j=1}^{3}(r_d-\tsup{r}_j)\prod_{k=1}^{7}(r_d-\tsup[1]{r}_k)^{-1}$. Note that the equation of motion \eqref{eq:drdphi_1} does not lead to any (hyper-)elliptic integrals. Hence, the inversion of such integrals cannot be performed using the existing methods. To simulate the orbits, we rely on numerical methods to derive the inversion. To proceed with this task, we consider a set of points $\big(r_i, \phi(r_i)=\phi_i\big)$ using the solution \eqref{eq:phi(r)_OFK}, and then we generate the numerical function $r(\phi)$ through interpolation. Based on this procedure, in Fig. \ref{fig:OFK}, we have plotted some examples of the OFK for two different scenarios.
First, we keep the impact parameter fixed and vary $\alpha$ to observe the effects of the nonminimal coupling (Fig. \ref{fig:OFK}(a)). As inferred from the diagram, the SBH exhibits the smallest deflection, as also indicated by its corresponding effective potential in Fig. \ref{fig:Veff}(a). Second, we fix the $\alpha$-parameter, which yields the effective potential in Fig. \ref{fig:Veff}(b), and vary the impact parameter to generate various orbits with different bending. As observed from diagram \ref{fig:OFK}(b), the closer the impact parameter approaches its critical value $b_p$, the more the trajectories exhibit attractive characteristics until they diverge and become repulsive in the vicinity of the effective potential's maximum.

\begin{figure}[h]
    \centering
    \includegraphics[width=7cm]{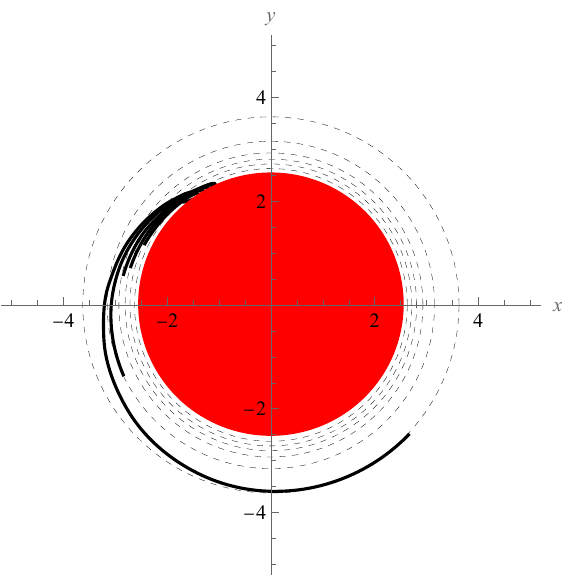} 
    \caption{The OSK plotted for $\phi_0=0$, $\delta=0.7$, and $\alpha=0.2$, in accordance with the configuration presented in Fig. \ref{fig:Veff}(b). Different impact parameters, ranging from $b_{d_1}$ to $b_{d_6}$, are displayed from outermost to innermost, as shown in Fig. \ref{fig:OFK}(b). The red disk represents the radius of $r_+$.
    The unit of length along the axes is $M$.}
    \label{fig:OSK}
\end{figure}

For the case of the OSK, the form of the solution is identical to that in Eq. \eqref{eq:phi(r)_OFK}; it is sufficient to perform the interchange $r_d\leftrightarrow r_f$. In Fig. \ref{fig:OSK}, several examples of the OSK have been plotted for a fixed $\alpha$ and a varying impact parameter.
%

%%%%%%%%%%%%%%%%%%%%
\subsubsection{Light deflection angle}

In fact, the light rays on the OFK are responsible for the famous phenomenon known as gravitational lensing, which arises from the deflection of light in the vicinity of massive objects. Gravitational lensing has emerged as a powerful tool for astrophysicists, offering a unique window into the universe. This phenomenon has facilitated numerous scientific breakthroughs since the advent of general relativity, and significant progress has been made in the past two decades in developing theoretical frameworks for calculating deflection angles and predicting lensing effects on the images of astrophysical objects (see, e.g., the seminal works in Refs. \cite{virbhadra_schwarzschild_2000,virbhadra_gravitational_2002,virbhadra_time_2008,virbhadra_relativistic_2009,virbhadra_distortions_2022,virbhadra_conservation_2024,virbhadra_compactness_2024}). 

\begin{figure}[h]
    \centering
    \includegraphics[width=7cm]{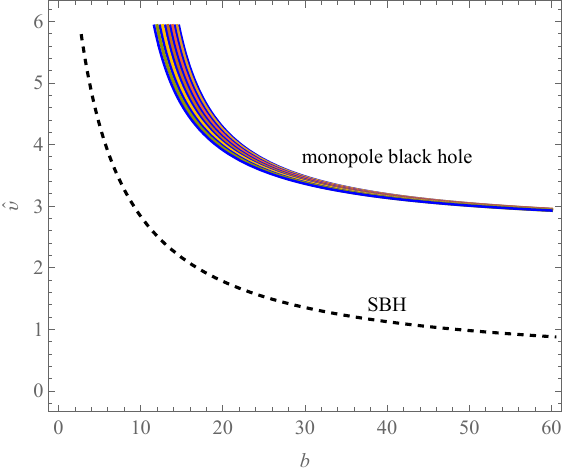} 
    \caption{The \( b \)-profile of the deflection angle plotted for \( \delta=0.7 \), illustrating the behavior of the deflection angle for the monopole black hole and the SBH. For the monopole black hole, the curves, from top to bottom, correspond to changes in the \( \alpha \)-parameter within the range \( -1<\alpha<1 \), with increments of 0.2.
    The unit of length along the axes is $M$.}
    \label{fig:hatupsilon}
\end{figure}

To calculate the bending angle of light, and considering that the spacetime under consideration is asymptotically flat, we can derive the deflection angle during the OFK at \( r_d \) using the equation of motion \eqref{eq:drdphi_1}:
\begin{equation}
\hat{\upsilon} = 2 \int_{r_d}^{\infty} \sqrt{\frac{\mathcal{P}_4(r)}{\mathcal{P}_8(r)}} \, \ed r - \pi.
    \label{eq:upsilonhat_0}
\end{equation}
Applying the same methods as used previously within this study, we obtain
\begin{equation}
\hat{\upsilon}=\frac{4b^2}{r_d^2}\,\tilde{F}_D^{(11)}\left(
1,\Bigl\{b_{\{\overline{1,3}\}}\Bigr\},\Bigl\{b_{\{\overline{4,10}\}}\Bigr\},\frac{3}{2};\Bigl\{\tsup{c}_{\{\overline{1,3}\}}\Bigr\},\Bigl\{\tsup{c}_{\{\overline{4,10}\}}\Bigr\}
\right)-\pi,
    \label{eq:upsilonhat_1}
\end{equation}
where $\Bigl\{b_{\{\overline{1,3}\}}\Bigr\}$ and $\Bigl\{b_{\{\overline{4,10}\}}\Bigr\}$ have been given in Eq. \eqref{eq:b13410}, and
\begin{subequations}
    \begin{align}
        & \Bigl\{\tsup{c}_{\{\overline{1,3}\}}\Bigr\}=\left\{\frac{  \tsup{r}_{\{\overline{1,3}\}} }{r_d}\right\}_3,\\
        & \Bigl\{\tsup{c}_{\{\overline{4,10}\}}\Bigr\}=\left\{\frac{  \tsup[1]{r}_{\{\overline{1,7}\}} }{r_d}\right\}_7,
    \end{align}
\end{subequations}
and 
\begin{equation}
\tilde{F}_D^{(n+1)}\left(
a,b_1,\dots,b_n,c;x_1,\dots,x_n
\right) = \frac{\Gamma(a)\Gamma(c-a)}{\Gamma(c)}\int_0^1 z^{a-1}(1-z)^{c-a-1}\prod_{j=1}^{n}(1-x_j z)^{-b_j}\ed z,
    \label{eq:tFD_def}
\end{equation}
is the Euler integral form of the definite $n$-parameter Lauricella hypergeometric function. In Fig. \ref{fig:hatupsilon}, the $b$-profile of the deflection angle $\hat{\upsilon}$ has been plotted for different values of the $\alpha$-parameter.
As inferred from the figure, the SBH provides a significantly lesser deflection angle than the monopole black hole. For the latter, an increase in the \( \alpha \)-parameter from negative to positive values results in a decrease in the deflection angle for the same \( b \). Therefore, we can expect the photons to experience more strongly deflected outgoing forward trajectories. This observation is consistent with what we see in Fig. \ref{fig:OFK}(a).

%%%%%%%%%%%%%%%%%%%%%%%%%%%%%%
\subsubsection{Critical trajectories}

\begin{figure}[h]
    \centering
    \includegraphics[width=5.3cm]{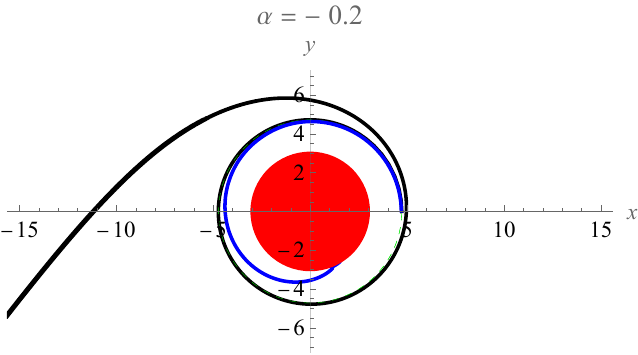} (a)
    \includegraphics[width=5.3cm]{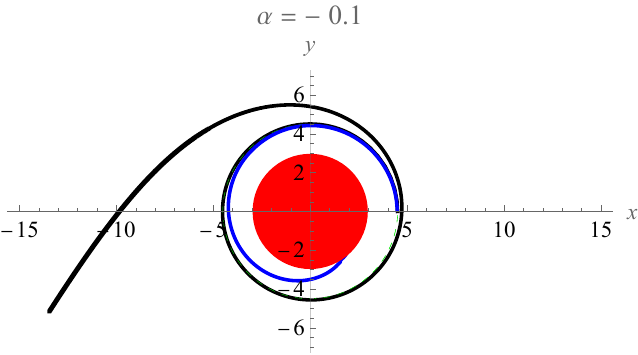} (b)
    \includegraphics[width=5.3cm]{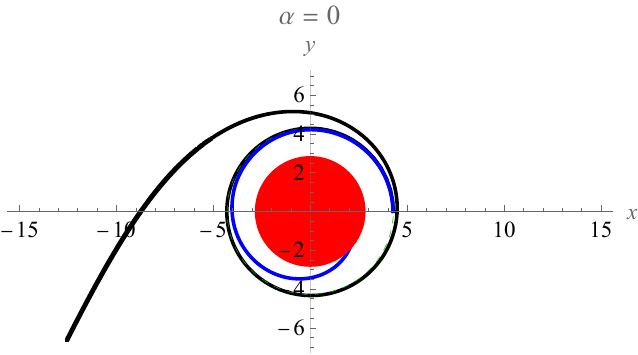} (c)
    \includegraphics[width=5.3cm]{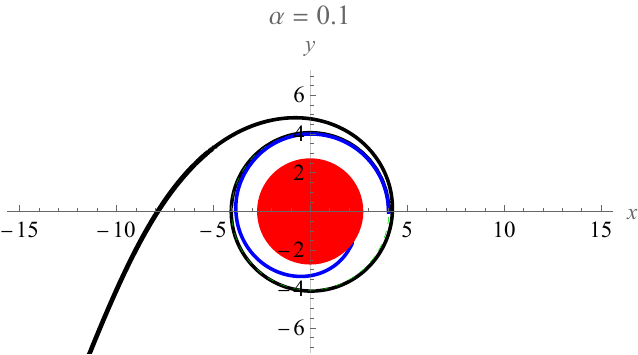} (d)
    \includegraphics[width=5.3cm]{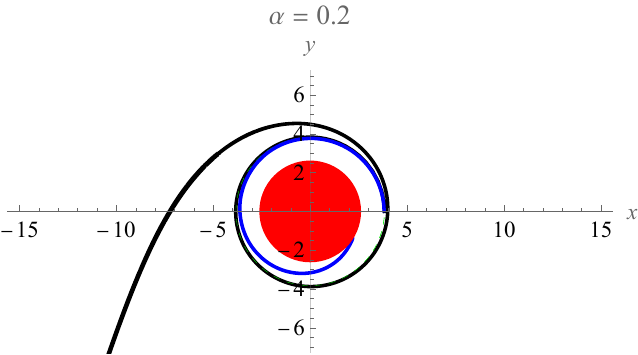} (e)
    \includegraphics[width=5.3cm]{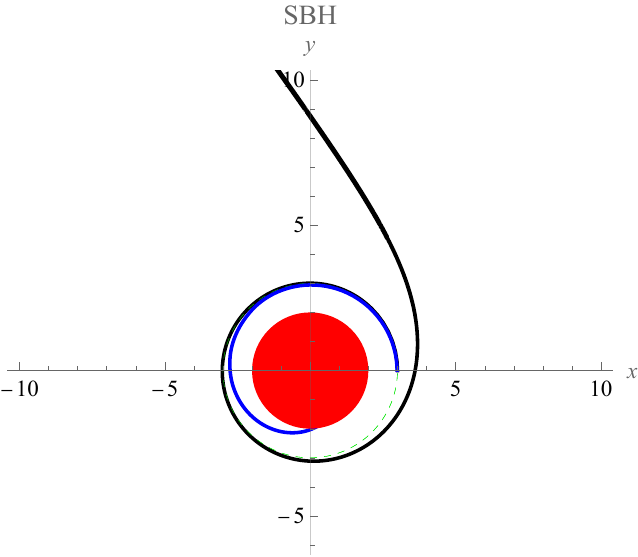} (f)
    \caption{The COFK (black curves) and COSK (blue curves) plotted for \( \delta = 0.7 \) and different values of the \( \alpha \)-parameter. Panel (f) indicates the critical orbits for the SBH. The dashed green circle in each diagram represents \( r_p \), which for each case is as follows: (a) \( r_p = 4.723 \) with \( b_p = 9.644 \), (b) \( r_p = 4.508 \) with \( b_p = 9.267 \), (c) \( r_p = 4.286 \) with \( b_p = 8.873 \), (d) \( r_p = 4.058 \) with \( b_p = 8.459 \), (e) \( r_p = 3.829 \) with \( b_p = 8.029 \), and (f) \( r_p = 3.0 \) with \( b_p = 5.196 \). The radius of the red disks corresponds to \( r_+ \).
    The unit of length along the axes is $M$.}
    \label{fig:critical}
\end{figure}

At the maximum of the effective potential, the two roots \( r_f \) and \( r_d \) of the polynomial \( \mathcal{P}_8 \) merge to form the degenerate root \( r_p \) (see Fig. \ref{fig:Veff}(b)). This scenario can be incorporated into the general solution in Eq. \eqref{eq:phi(r)_OFK}, which reduces the 10-parameter function to a 9-parameter Lauricella hypergeometric function. Considering this, and following the procedures established earlier, we have illustrated the COFK and COSK for different values of the \( \alpha \)-parameter in Fig. \ref{fig:critical}.
%

%%%%%%%%%%%%%%%%%%%%%%%%%%%%%%
\subsubsection{The capture zone}

\begin{figure}[h]
    \centering
    \includegraphics[width=7cm]{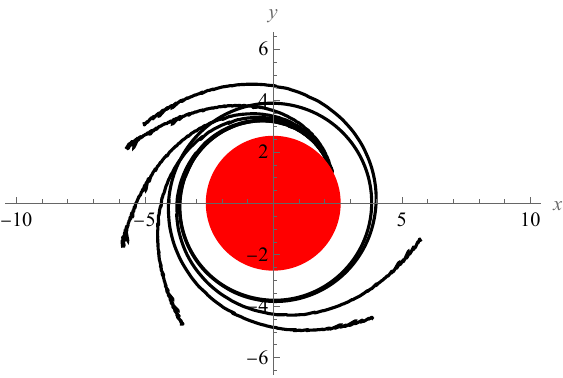} 
    \caption{The capture zone plotted for \( \alpha = 0.2 \) and \( \delta = 0.7 \), in accordance with the configuration in Fig. \ref{fig:Veff}(b). The range of the implemented impact parameter varies from \( b = b_p - 0.001 \) to \( b = b_p - 1.5 \).
    The radius of the red disks is \( r_+ \).
    The unit of length along the axes is $M$.}
    \label{fig:capture}
\end{figure}

Once \( b < b_p \), the roots of the polynomial equation \( \mathcal{P}_8(r) = 0 \) are either negative or complex conjugate. Consequently, the photons do not encounter any turning points, leading to a direct infall onto the black hole's event horizon. By applying this condition to the solution in Eq. \eqref{eq:phi(r)_OFK}, we have demonstrated the capture orbits for different values of the impact parameter in Fig. \ref{fig:capture}, corresponding to the capture zone, in accordance with the effective potential configuration shown in Fig. \ref{fig:Veff}(b).
In fact, as the impact parameter approaches \( b_p \) from below, the trajectories become increasingly spiral before infalling onto the black hole. As observed from Fig. \ref{fig:capture}, all the orbits converge into a single curve prior to their infall onto the event horizon, and collectively, all the captured orbits form the capture zone. \\

Thus far, we have investigated, both analytically and numerically, the geodesic structure of light-like trajectories around the monopole black hole. Consequently, the general objective of this study has been addressed in full detail. However, before concluding this discussion, it is worthwhile to provide some astrophysical insights into the black hole's structure. Therefore, based on our findings in this section, the next section will present a brief study on the confinement of the black hole parameters, based on the recent EHT observations.

%%%%%%%%%%%%%%%%%%sect.IV
\section{Observational constraints on the black hole parameters}\label{sec:EHTconst.}

\begin{figure}[h]
    \centering
    \includegraphics[width=7cm]{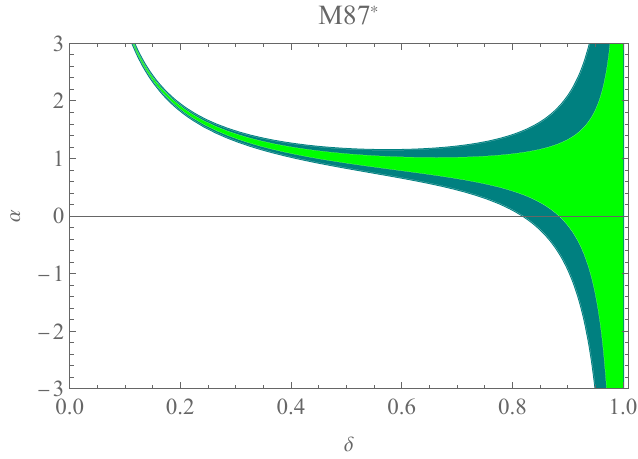} (a)\qquad
    \includegraphics[width=7cm]{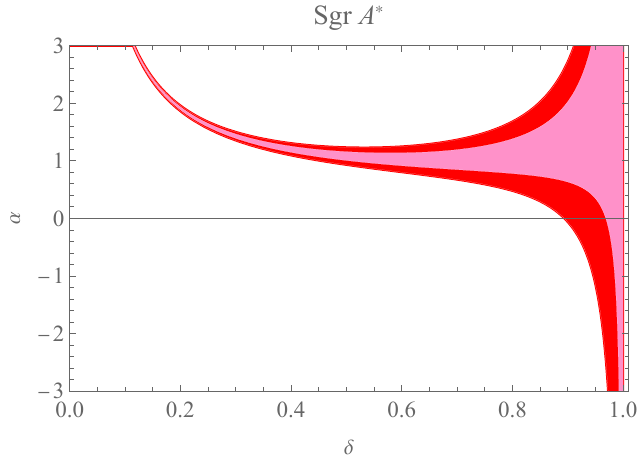} (b)
    \caption{Constraints on $\alpha$ and $\delta$ within the observed shadow diameters of M87* and Sgr A* at $1\sigma$ (lighter colors) and $2\sigma$ (darker colors) uncertainties.}
    \label{fig:constraints}
\end{figure}

As previously mentioned, the photon sphere is characterized by photons on unstable orbits. These photons, influenced by the gravitational field of black holes, will either spiral into the event horizon (COSK) or escape (COFK). Those that manage to escape create a luminous photon ring that surrounds the black hole's shadow \cite{Synge:1966,Cunningham:1972,Bardeen:1973a,Luminet:1979}. Thus, the photons on unstable orbits delineate the boundary of the black hole shadow as observed from a distance. To facilitate comparisons with the EHT observations, we can compute the theoretical diameter of the black hole shadow, which is given by 
$d_{\mathrm{sh}}^{\mathrm{theo}} = 2R_{s} = 2b_p$, where \( b_p \) is defined in Eq. \eqref{eq:bp}. For the actual diameter of the shadows observed in the recent EHT images of M87* and Sgr A*, we utilize the relation \cite{Bambi:2019tjh}
\begin{equation}
d_\text{sh} = \frac{D \theta_*}{\gamma M_\odot}, \label{eq:dsh}
\end{equation}
where \( D \) is the distance to the observer (in parsecs), and \( \gamma \) represents the mass ratio of the black hole to the Sun. For M87*, the mass ratio is \( \gamma = (6.5 \pm 0.90) \times 10^9 \) at a distance \( D = 16.8\,\mathrm{Mpc} \) \cite{Akiyama:2019}, while for Sgr A*, it is \( \gamma = (4.3 \pm 0.013) \times 10^6 \) at \( D = 8.127\,\mathrm{kpc} \) \cite{Akiyama:2022}. The angular diameter of the shadow, denoted \( \theta_* \), is measured as \( \theta_* = (42 \pm 3) \,\mathrm{\mu as} \) for M87* and \( \theta_* = (48.7 \pm 7) \,\mathrm{\mu as} \) for Sgr A*. Using these measurements, we can derive the shadow diameters as \( d_{\mathrm{sh}}^{\mathrm{M87*}} = (11 \pm 1.5)M \) and \( d_{\mathrm{sh}}^{\mathrm{SgrA*}} = (9.5 \pm 1.4)M \), which are represented in Fig. \ref{fig:constraints} along with their \( 1\sigma \) and \( 2\sigma \) uncertainties.
The diagrams indicate that an increase in \( \delta \) expands the permissible range for the \( \alpha \)-parameter based on the observational data. As \( \delta \) approaches 1 (the SBH), the reliable range for \( \alpha \) tends toward infinity, which is expected since the SBH is independent of this parameter. Conversely, smaller values of \( \delta \) correlate with a narrower acceptable range for \( \alpha \), typically encompassing larger values. However, since \( \alpha \) is expected to be small, such values can be overlooked. Consequently, from the diagrams, we find the most reliable ranges for the black hole parameters to be \( 0.5 \lesssim \delta \leq 1 \) and \( -0.5 \lesssim \alpha \lesssim 0.5 \). These ranges align with those presented in Ref. \cite{carames_nonminimal_2023}, based on the light deflection angle, supporting the criteria \( \delta \approx 1 \) and \( 0 < \alpha \ll  1\).

%%%%%%%%%%%%%%%%%%%
\section{Summary and conclusions}\label{sec:conclusions}

In this work, we performed an analytical investigation of null geodesics around a black hole with a weakly coupled global monopole charge. We first reviewed the black hole spacetime in the presence of a global monopole, showing that the positiveness and negativeness of the coupling constant $\alpha$ introduces significant modifications to the event horizon structure, with the black hole exhibiting two horizons for $\alpha < 0$ and one horizon for $\alpha \geq 0$. These features were important for understanding the causal structure in this spacetime. Using the Lagrangian formalism, we derived and analyzed the behavior of both radial and angular null trajectories, focusing on the effects of the nonminimal coupling parameter $\alpha$ and the deficit angle $\delta$. The effective potential governing photon motion was found to be sensitive to the nonminimal coupling. In particular, we observed that positive $\alpha$ leads to a higher effective potential barrier, resulting in stronger photon deflection near the black hole, while negative $\alpha$ reduces the barrier. This distinction is crucial for understanding the deflection angles and shadow formation around such black holes. We categorized photon orbits based on their impact parameters and demonstrated how different values of $\alpha$ influence the types of photon trajectories, including deflecting and critical orbits. We argued that the solutions to the equations of motion cannot be expressed in terms of elliptic integrals. Instead, we presented our solutions using Lauricella incomplete hypergeometric functions of different orders. These solutions were provided for both radial and angular photon motion, and the inversion of the integrals was performed numerically to simulate the orbits. Furthermore, to calculate the light deflection angle, a finite Lauricella function was employed, and the results were compared to the deflection angle of the SBH. Additionally, we compared our theoretical results with the EHT observations of M87* and Sgr A*, using the shadow diameter as a key observable. The calculated shadow diameter allowed us to place constraints on the parameters $\alpha$ and $\delta$, with the most reliable ranges being $0.5 \lesssim \delta \leq 1$ and $-0.5 \lesssim \alpha \lesssim 0.5 $. As $\delta$ approaches 1, the black hole approaches the Schwarzschild limit, leading to a broader allowed range for $\alpha$. Our findings could provide new insights into the geodesic structure of black holes with global monopole charges and offer a framework for constraining such models with observational data. The results suggest that future observations of black hole shadows, particularly with higher precision, can further constrain the parameters $\alpha$ and $\delta$, shedding light on the role of global monopoles in astrophysical black holes.

%\textcolor{Orange}{\textbf{$\bullet$ Mohsen está trabajando aquí$\longrightarrow$}}

%%%%%%%%%%%%
\begin{acknowledgments}
M.F. is supported by Universidad Central de Chile through project No. PDUCEN20240008.  J.R.V. is partially supported by Centro de F\'isica Teórica de Valparaíso (CeFiTeV). 
\end{acknowledgments}
%%%%%%%%%%%%

%\section{References}
%%%%%%%%%%%%%References
\bibliographystyle{ieeetr}
\bibliography{biblio_v1}

\begin{thebibliography}{10}

\bibitem{shellard_vilenkin}
A.~Vilenkin and E.~Shellard, {\em Cosmic Strings and Other Topological Defects}.
\newblock Cambridge Monographs on Mathematical Physics, Cambridge University Press, 1994.

\bibitem{Kibble_1976}
T.~W.~B. Kibble, ``Topology of cosmic domains and strings,'' {\em Journal of Physics A: Mathematical and General}, vol.~9, p.~1387, aug 1976.

\bibitem{Barriola-Vilenkin:1989}
M.~Barriola and A.~Vilenkin, ``Gravitational field of a global monopole,'' {\em Phys. Rev. Lett.}, vol.~63, pp.~341--343, Jul 1989.

\bibitem{Harari-Lousto}
D.~Harari and C.~Loust\'o, ``Repulsive gravitational effects of global monopoles,'' {\em Phys. Rev. D}, vol.~42, pp.~2626--2631, Oct 1990.

\bibitem{CRomero}
A.~Barros and C.~Romero, ``Global monopoles in brans-dicke theory of gravity,'' {\em Phys. Rev. D}, vol.~56, pp.~6688--6691, Nov 1997.

\bibitem{Liu_2009}
D.-J. Liu, Y.-L. Zhang, and X.-Z. Li, ``A self-gravitating dirac–born–infeld global monopole,'' {\em The European Physical Journal C}, vol.~60, p.~495–500, Feb. 2009.

\bibitem{Carames_2011}
T.~R.~P. Caramês, E.~R. Bezerra~de Mello, and M.~E.~X. Guimarães, ``Gravitational field of a global monopole in a modified gravity,'' {\em International Journal of Modern Physics: Conference Series}, vol.~03, p.~446–454, Jan. 2011.

\bibitem{Carames_2017}
T.~R.~P. Caramês, J.~C. Fabris, E.~R. Bezerra~de Mello, and H.~Belich, ``$f(\textrm{R})$ global monopole revisited,'' {\em The European Physical Journal C}, vol.~77, July 2017.

\bibitem{Lambaga_2018}
R.~D. Lambaga and H.~S. Ramadhan, ``Gravitational field of global monopole within the eddington-inspired born-infeld theory of gravity,'' {\em The European Physical Journal C}, vol.~78, May 2018.

\bibitem{Nascimento_2019}
J.~R. Nascimento, G.~J. Olmo, P.~J. Porf\'{\i}rio, A.~Y. Petrov, and A.~R. Soares, ``Global monopole in palatini $f(\mathcal{R})$ gravity,'' {\em Phys. Rev. D}, vol.~99, p.~064053, Mar 2019.

\bibitem{gusmann_scattering_2021}
A.~Gußmann, ``Scattering of axial gravitational wave pulses by monopole black holes and {QNMs}: a semianalytic approach,'' {\em Classical and Quantum Gravity}, vol.~38, p.~035008, Feb. 2021.

\bibitem{GBgravity-2023}
N.~Chatzifotis, N.~E. Mavromatos, and D.~P. Theodosopoulos, ``Global monopoles in the extended gauss-bonnet gravity,'' {\em Phys. Rev. D}, vol.~107, p.~085014, Apr 2023.

\bibitem{carames_nonminimal_2023}
T.~R.~P. Caramês, ``Nonminimal global monopole,'' {\em Physical Review D}, vol.~108, p.~084002, Oct. 2023.

\bibitem{1920RSPTA.220..291D}
F.~W. {Dyson}, A.~S. {Eddington}, and C.~{Davidson}, ``{A Determination of the Deflection of Light by the Sun's Gravitational Field, from Observations Made at the Total Eclipse of May 29, 1919},'' {\em Philosophical Transactions of the Royal Society of London Series A}, vol.~220, pp.~291--333, Jan. 1920.

\bibitem{RevModPhys.19.361}
G.~M. Clemence, ``The relativity effect in planetary motions,'' {\em Rev. Mod. Phys.}, vol.~19, pp.~361--364, Oct 1947.

\bibitem{1930JaJAG...8...67H}
Y.~{Hagihara}, ``{Theory of the Relativistic Trajeetories in a Gravitational Field of Schwarzschild},'' {\em Japanese Journal of Astronomy and Geophysics}, vol.~8, p.~67, Jan. 1930.

\bibitem{jacobi_2013}
C.~G.~J. Jacobi, {\em C. G. J. Jacobi's Gesammelte Werke: Herausgegeben auf Veranlassung der königlich preussischen Akademie der Wissenschaften}.
\newblock Cambridge Library Collection - Mathematics, Cambridge University Press, 2013.

\bibitem{abel_2012}
N.~H. Abel, {\em Oeuvres complètes de Niels Henrik Abel: Nouvelle édition}, vol.~2 of {\em Cambridge Library Collection - Mathematics}.
\newblock Cambridge University Press.

\bibitem{Riemann:1857}
B.~Riemann, ``Theorie der {Abel}'schen {Functionen}.,'' {\em Journal für die reine und angewandte Mathematik (Crelles Journal)}, vol.~1857, pp.~115--155, July 1857.

\bibitem{Riemann+1866+161+172}
B.~Riemann, ``Ueber das verschwinden der $\vartheta$-functionen.,'' {\em Journal für die reine und angewandte Mathematik (Crelles Journal)}, vol.~1866, no.~65, pp.~161--172, 1866.

\bibitem{Weierstrass+1854+289+306}
C.~Weierstrass, ``Zur theorie der abelschen functionen.,'' {\em Journal für die reine und angewandte Mathematik (Crelles Journal)}, vol.~1854, no.~47, pp.~289--306, 1854.

\bibitem{kraniotis_general_2002}
G.~V. Kraniotis and S.~B. Whitehouse, ``General relativity, the cosmological constant and modular forms,'' {\em Classical and Quantum Gravity}, vol.~19, pp.~5073--5100, Oct. 2002.

\bibitem{kraniotis_compact_2003}
G.~V. Kraniotis and S.~B. Whitehouse, ``Compact calculation of the perihelion precession of {Mercury} in general relativity, the cosmological constant and {Jacobi}'s inversion problem,'' {\em Classical and Quantum Gravity}, vol.~20, pp.~4817--4835, Nov. 2003.

\bibitem{kraniotis_precise_2004}
G.~V. Kraniotis, ``Precise relativistic orbits in {Kerr} and {Kerr}–(anti) de {Sitter} spacetimes,'' {\em Classical and Quantum Gravity}, vol.~21, pp.~4743--4769, Oct. 2004.

\bibitem{kraniotis_frame_2005}
G.~V. Kraniotis, ``Frame dragging and bending of light in {Kerr} and {Kerr}–(anti) de {Sitter} spacetimes,'' {\em Classical and Quantum Gravity}, vol.~22, pp.~4391--4424, Nov. 2005.

\bibitem{cruz_geodesic_2005}
N.~Cruz, M.~Olivares, and J.~R. Villanueva, ``The geodesic structure of the {Schwarzschild} anti-de {Sitter} black hole,'' {\em Classical and Quantum Gravity}, vol.~22, pp.~1167--1190, Mar. 2005.

\bibitem{kraniotis_periapsis_2007}
G.~V. Kraniotis, ``Periapsis and gravitomagnetic precessions of stellar orbits in {Kerr} and {Kerr}–de {Sitter} black hole spacetimes,'' {\em Classical and Quantum Gravity}, vol.~24, pp.~1775--1808, Apr. 2007.

\bibitem{hackmann_complete_2008}
E.~Hackmann and C.~Lämmerzahl, ``Complete {Analytic} {Solution} of the {Geodesic} {Equation} in {Schwarzschild}–({Anti}-)de {Sitter} {Spacetimes},'' {\em Physical Review Letters}, vol.~100, p.~171101, May 2008.

\bibitem{hackmann_geodesic_2008}
E.~Hackmann and C.~Lämmerzahl, ``Geodesic equation in {Schwarzschild}-(anti-)de {Sitter} space-times: {Analytical} solutions and applications,'' {\em Physical Review D}, vol.~78, p.~024035, July 2008.

\bibitem{hackmann_analytic_2009}
E.~Hackmann, V.~Kagramanova, J.~Kunz, and C.~Lämmerzahl, ``Analytic solutions of the geodesic equation in axially symmetric space-times,'' {\em EPL (Europhysics Letters)}, vol.~88, p.~30008, Nov. 2009.

\bibitem{hackmann_complete_2010}
E.~Hackmann, B.~Hartmann, C.~Lämmerzahl, and P.~Sirimachan, ``Complete set of solutions of the geodesic equation in the space-time of a {Schwarzschild} black hole pierced by a cosmic string,'' {\em Physical Review D}, vol.~81, p.~064016, Mar. 2010.

\bibitem{olivares_motion_2011}
M.~Olivares, J.~Saavedra, C.~Leiva, and J.~R. Villanueva, ``{Motion of charged particles on the Reissner–Nordstr\"om (anti)-de Sitter black hole spacetime},'' {\em Modern Physics Letters A}, vol.~26, pp.~2923--2950, Dec. 2011.

\bibitem{kraniotis_precise_2011}
G.~V. Kraniotis, ``Precise analytic treatment of {Kerr} and {Kerr}-(anti) de {Sitter} black holes as gravitational lenses,'' {\em Classical and Quantum Gravity}, vol.~28, p.~085021, Apr. 2011.

\bibitem{cruz_geodesic_2013}
N.~Cruz, M.~Olivares, and J.~R. Villanueva, ``Geodesic structure of {Lifshitz} black holes in 2+1 dimensions,'' {\em The European Physical Journal C}, vol.~73, p.~2485, July 2013.

\bibitem{villanueva_photons_2013}
J.~R. Villanueva, J.~Saavedra, M.~Olivares, and N.~Cruz, ``Photons motion in charged {Anti}-de {Sitter} black holes,'' {\em Astrophysics and Space Science}, vol.~344, pp.~437--446, Apr. 2013.

\bibitem{kraniotis_gravitational_2014}
G.~V. Kraniotis, ``Gravitational lensing and frame dragging of light in the {Kerr}–{Newman} and the {Kerr}–{Newman} (anti) de {Sitter} black hole spacetimes,'' {\em General Relativity and Gravitation}, vol.~46, p.~1818, Nov. 2014.

\bibitem{soroushfar_analytical_2015}
S.~Soroushfar, R.~Saffari, J.~Kunz, and C.~Lämmerzahl, ``Analytical solutions of the geodesic equation in the spacetime of a black hole in f ( {R} ) gravity,'' {\em Physical Review D}, vol.~92, p.~044010, Aug. 2015.

\bibitem{soroushfar_detailed_2016}
S.~Soroushfar, R.~Saffari, S.~Kazempour, S.~Grunau, and J.~Kunz, ``Detailed study of geodesics in the {Kerr}-{Newman}-({A}){dS} spacetime and the rotating charged black hole spacetime in f ( {R} ) gravity,'' {\em Physical Review D}, vol.~94, p.~024052, July 2016.

\bibitem{hoseini_analytic_2016}
B.~Hoseini, R.~Saffari, S.~Soroushfar, S.~Grunau, and J.~Kunz, ``Analytic treatment of complete geodesics in a static cylindrically symmetric conformal spacetime,'' {\em Physical Review D}, vol.~94, p.~044021, Aug. 2016.

\bibitem{hoseini_study_2017}
B.~Hoseini, R.~Saffari, and S.~Soroushfar, ``Study of the geodesic equations of a spherical symmetric spacetime in conformal {Weyl} gravity,'' {\em Classical and Quantum Gravity}, vol.~34, p.~055004, Mar. 2017.

\bibitem{fathi_motion_2020}
M.~Fathi, M.~Kariminezhaddahka, M.~Olivares, and J.~R. Villanueva, ``Motion of massive particles around a charged {Weyl} black hole and the geodetic precession of orbiting gyroscopes,'' {\em The European Physical Journal C}, vol.~80, p.~377, May 2020.

\bibitem{fathi_classical_2020}
M.~Fathi, M.~Olivares, and J.~R. Villanueva, ``Classical tests on a charged {Weyl} black hole: bending of light, {Shapiro} delay and {Sagnac} effect,'' {\em The European Physical Journal C}, vol.~80, p.~51, Jan. 2020.

\bibitem{fathi_gravitational_2021}
M.~Fathi, M.~Olivares, and J.~R. Villanueva, ``Gravitational {Rutherford} scattering of electrically charged particles from a charged {Weyl} black hole,'' {\em The European Physical Journal Plus}, vol.~136, p.~420, Apr. 2021.

\bibitem{gonzalez_null_2021}
P.~A. González, M.~Olivares, Y.~Vásquez, and J.~R. Villanueva, ``Null geodesics in five-dimensional {Reissner}–{Nordström} anti-de {Sitter} black holes,'' {\em The European Physical Journal C}, vol.~81, p.~236, Mar. 2021.

\bibitem{fathi_analytical_2021}
M.~Fathi, M.~Olivares, and J.~R. Villanueva, ``Analytical study of light ray trajectories in {Kerr} spacetime in the presence of an inhomogeneous anisotropic plasma,'' {\em The European Physical Journal C}, vol.~81, p.~987, Nov. 2021.

\bibitem{kraniotis_gravitational_2021}
G.~V. Kraniotis, ``Gravitational redshift/blueshift of light emitted by geodesic test particles, frame-dragging and pericentre-shift effects, in the {Kerr}–{Newman}–de {Sitter} and {Kerr}–{Newman} black hole geometries,'' {\em The European Physical Journal C}, vol.~81, p.~147, Feb. 2021.

\bibitem{fathi_study_2022}
M.~Fathi, M.~Olivares, and J.~R. Villanueva, ``Study of null and time-like geodesics in the exterior of a {Schwarzschild} black hole with quintessence and cloud of strings,'' {\em The European Physical Journal C}, vol.~82, p.~629, July 2022.

\bibitem{soroushfar_analytical_2022}
S.~Soroushfar and M.~Afrooz, ``Analytical solutions of the geodesic equation in the space-time of a black hole surrounded by perfect fluid in {Rastall} theory,'' {\em Indian Journal of Physics}, vol.~96, pp.~593--607, Feb. 2022.

\bibitem{battista_geodesic_2022}
E.~Battista and G.~Esposito, ``Geodesic motion in {Euclidean} {Schwarzschild} geometry,'' {\em The European Physical Journal C}, vol.~82, p.~1088, Dec. 2022.

\bibitem{fathi_spherical_2023}
M.~Fathi, M.~Olivares, and J.~R. Villanueva, ``Spherical photon orbits around a rotating black hole with quintessence and cloud of strings,'' {\em The European Physical Journal Plus}, vol.~138, p.~7, Jan. 2023.

\bibitem{fathi_analytical_2023}
M.~Fathi, ``Analytical study of particle geodesics around a scale-dependent de {Sitter} black hole,'' {\em Annals of Physics}, vol.~457, p.~169401, Oct. 2023.

\bibitem{Bertolami:2007}
O.~Bertolami, C.~G. B\"ohmer, T.~Harko, and F.~S.~N. Lobo, ``Extra force in $f(\textrm{R})$ modified theories of gravity,'' {\em Phys. Rev. D}, vol.~75, p.~104016, May 2007.

\bibitem{PhysRevD.105.024020}
S.~B. Fisher and E.~D. Carlson, ``Nuclear limits on nonminimally coupled gravity,'' {\em Phys. Rev. D}, vol.~105, p.~024020, Jan 2022.

\bibitem{PhysRevD.105.044048}
R.~March, O.~Bertolami, M.~Muccino, C.~Gomes, and S.~Dell'Agnello, ``Cassini and extra force constraints to nonminimally coupled gravity with a screening mechanism,'' {\em Phys. Rev. D}, vol.~105, p.~044048, Feb 2022.

\bibitem{Misner:1973}
C.~W. Misner, K.~S. Thorne, and J.~A. Wheeler, {\em Gravitation}.
\newblock Princeton University Press, 2017.

\bibitem{Chandrasekhar:2002}
S.~Chandrasekhar, {\em The mathematical theory of black holes}.
\newblock Oxford classic texts in the physical sciences, Oxford University Press, 2002.

\bibitem{ryder_2009}
L.~Ryder, {\em Introduction to General Relativity}.
\newblock Cambridge University Press, 2009.

\bibitem{virbhadra_schwarzschild_2000}
K.~S. Virbhadra and G.~F.~R. Ellis, ``Schwarzschild black hole lensing,'' {\em Physical Review D}, vol.~62, p.~084003, Sept. 2000.

\bibitem{virbhadra_gravitational_2002}
K.~S. Virbhadra and G.~F.~R. Ellis, ``Gravitational lensing by naked singularities,'' {\em Physical Review D}, vol.~65, p.~103004, May 2002.

\bibitem{virbhadra_time_2008}
K.~S. Virbhadra and C.~R. Keeton, ``Time delay and magnification centroid due to gravitational lensing by black holes and naked singularities,'' {\em Physical Review D}, vol.~77, p.~124014, June 2008.

\bibitem{virbhadra_relativistic_2009}
K.~S. Virbhadra, ``Relativistic images of {Schwarzschild} black hole lensing,'' {\em Physical Review D}, vol.~79, p.~083004, Apr. 2009.

\bibitem{virbhadra_distortions_2022}
K.~Virbhadra, ``Distortions of images of {Schwarzschild} lensing,'' {\em Physical Review D}, vol.~106, p.~064038, Sept. 2022.

\bibitem{virbhadra_conservation_2024}
K.~Virbhadra, ``Conservation of distortion of gravitationally lensed images,'' {\em Physical Review D}, vol.~109, p.~124004, June 2024.

\bibitem{virbhadra_compactness_2024}
K.~Virbhadra, ``Compactness of supermassive dark objects at galactic centers,'' {\em Canadian Journal of Physics}, pp.~cjp--2023--0313, June 2024.

\bibitem{Synge:1966}
J.~L. Synge, ``{The Escape of Photons from Gravitationally Intense Stars},'' {\em Mon. Not. Roy. Astron. Soc.}, vol.~131, pp.~463--466, 02 1966.

\bibitem{Cunningham:1972}
C.~T. {Cunningham} and J.~M. {Bardeen}, ``{The Optical Appearance of a Star Orbiting an Extreme Kerr Black Hole},'' {\em Astrophys. J. Letters}, vol.~173, p.~L137, May 1972.

\bibitem{Bardeen:1973a}
J.~M. {Bardeen}, W.~H. {Press}, and S.~A. {Teukolsky}, ``{Rotating Black Holes: Locally Nonrotating Frames, Energy Extraction, and Scalar Synchrotron Radiation},'' {\em Astrophys. J}, vol.~178, pp.~347--370, Dec. 1972.

\bibitem{Luminet:1979}
J.~P. {Luminet}, ``{Image of a spherical black hole with thin accretion disk.},'' {\em A\&A}, vol.~75, pp.~228--235, May 1979.

\bibitem{Bambi:2019tjh}
C.~Bambi, K.~Freese, S.~Vagnozzi, and L.~Visinelli, ``{Testing the rotational nature of the supermassive object M87* from the circularity and size of its first image},'' {\em Phys. Rev. D}, vol.~100, no.~4, p.~044057, 2019.

\bibitem{Akiyama:2019}
K.~Akiyama {\em et~al.}, ``{First M87 Event Horizon Telescope Results. IV. Imaging the Central Supermassive Black Hole},'' {\em Astrophys. J. Lett.}, vol.~875, no.~1, p.~L4, 2019.

\bibitem{Akiyama:2022}
K.~Akiyama {\em et~al.}, ``{First Sagittarius A{$^{*}$} Event Horizon Telescope results. I. The shadow of the supermassive black hole in the center of the Milky Way.},'' {\em Astrophys. J. Lett.}, vol.~930, no.~2, p.~L12, 2022.

\end{thebibliography}
%\bibliography{DymnikovaTestEnero2024}

\end{document}